\documentclass[conference]{IEEEtran}
\IEEEoverridecommandlockouts
\usepackage{cite}
\usepackage{amsmath,amssymb,amsfonts}
\usepackage{algorithmic}
\usepackage{graphicx}
\usepackage{textcomp}
\usepackage{xcolor}

\usepackage[ruled, boxed,vlined,linesnumbered]{algorithm2e}
\usepackage{subfig} 
\usepackage{enumitem} 
\usepackage{multirow} 
\usepackage{array}
\usepackage{hyperref}
\usepackage{tikz}

\usepackage{balance}

\newcommand{\name}{SeSeMI}
\newcommand{\ksname} {KeyService}
\newcommand{\rtname}{SeMIRT} 
\newcommand{\fpname}{FnPacker} 
\newcommand{\ignore}[1]{}

\def\BibTeX{{\rm B\kern-.05em{\sc i\kern-.025em b}\kern-.08em
    T\kern-.1667em\lower.7ex\hbox{E}\kern-.125emX}}
\begin{document}

\title{\name{}: Secure Serverless Model Inference on Sensitive Data}

\author{
    \IEEEauthorblockN{Guoyu Hu\IEEEauthorrefmark{1}, Yuncheng Wu\IEEEauthorrefmark{2}, Gang Chen\IEEEauthorrefmark{3},  Tien Tuan Anh Dinh\IEEEauthorrefmark{4}, Beng Chin Ooi\IEEEauthorrefmark{1}}
    \IEEEauthorblockA{\IEEEauthorrefmark{1}National University of Singapore,  \IEEEauthorrefmark{2}Renmin University of China, \IEEEauthorrefmark{3}Zhejiang University, \IEEEauthorrefmark{4}Deakin University
    \\ guoyu.hu@u.nus.edu, ooibc@comp.nus.edu.sg,  wuyuncheng@ruc.edu.cn, cg@zju.edu.cn, anh.dinh@deakin.edu.au}
}

\maketitle

\begin{abstract}
Model inference systems are essential for implementing end-to-end data analytics pipelines that deliver the benefits of machine learning models to users. Existing cloud-based model inference systems are costly, not easy to scale, and must be trusted in handling the models and user request data. Serverless computing presents a new opportunity, as it provides elasticity and fine-grained pricing.

Our goal is to design a serverless model inference system that protects models and user request data from untrusted cloud providers. It offers high performance and low cost, while requiring no intrusive changes to the current serverless platforms. To realize our goal, we leverage trusted hardware. We identify and address three challenges in using trusted hardware for serverless model inference. These challenges arise from the high-level abstraction of serverless computing, the performance overhead of trusted hardware, and the characteristics of model inference workloads.

We present \name{}, a \underline{se}cure, efficient, and cost-effective \underline{se}rverless \underline{m}odel \underline{i}nference system.  It adds three novel features non-intrusively to the existing serverless infrastructure and nothing else. The first feature is a key service that establishes secure channels between the user and the serverless instances, which also provides access control to models and users’ data. The second is an enclave runtime that allows one enclave to process multiple concurrent requests. The final feature is a model packer that allows multiple models to be executed by one serverless instance. We build \name{} on top of Apache OpenWhisk, and conduct extensive experiments with three popular machine learning models. The results show that \name{} achieves low latency and low cost at scale for realistic workloads.

\end{abstract}


\section{Introduction}
\label{sec:introduction}

Machine learning (ML) model inference is an essential step in modern end-to-end data analytics pipelines.
The data to be processed undergoes typical preprocessing stages, such as cleaning, transformation, and feature extraction, and then is fed to ML models for inference.
Most cloud vendors offer services that manage the model serving on behalf of the user (or model owner). Examples include Amazon SageMaker~\footnote{\url{https://aws.amazon.com/sagemaker/}} and Google Vertex AI~\footnote{\url{https://cloud.google.com/vertex-ai}}. These services let the model owner deploy their models to pre-specified cloud instances and scale the
number of instances based on projected workloads.

\begin{figure}[t]
    \centering
    \includegraphics[width=0.48\textwidth]{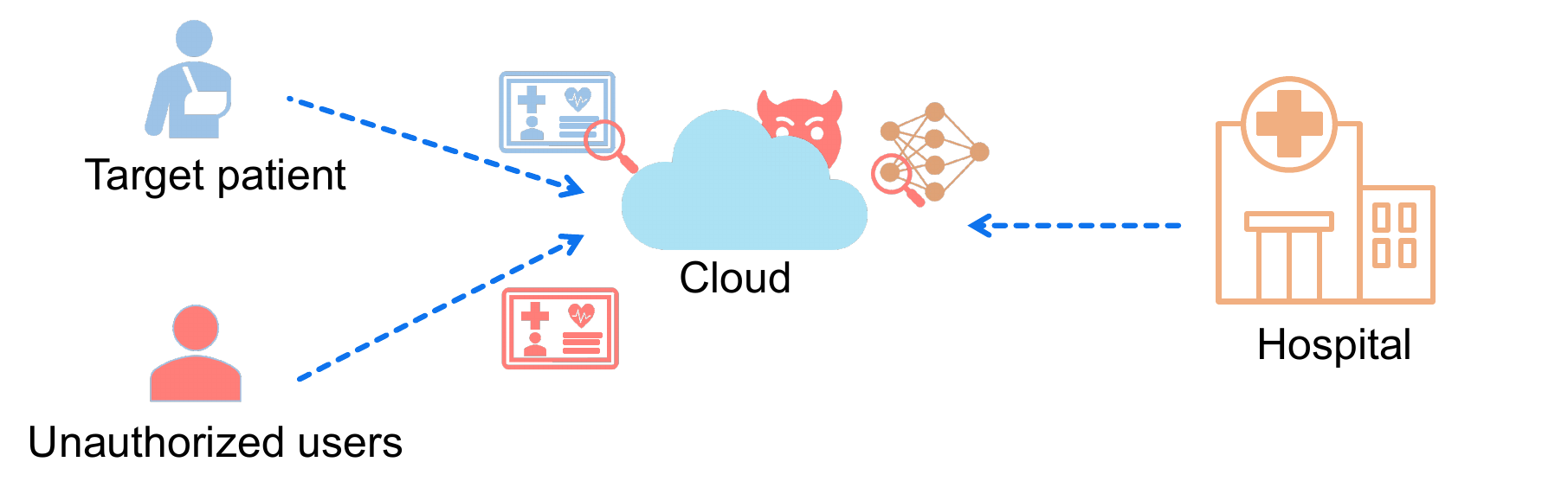}
    \vspace{-2mm}
    \caption{Security issue of cloud model inference services.}
    \label{fig:security-concern}
\end{figure}

Existing cloud-based model inference systems have two main limitations.
First, they are costly, as the model owner needs to over-provision resources for peak loads, and they are slow to scale, as they need to start new instances when scaling up~\cite{mark, our-work}. 
Second, they offer no privacy and access control.
Figure~\ref{fig:security-concern} depicts an example in which a hospital develops a disease prediction model based on its patients' electronic health records (EHRs) and uploads it to the cloud. The cloud deploys the model over multiple virtual machine instances. 
It accepts requests from authorized patients and doctors and executes model inference over their data. In this example, there is no privacy protection, because both the request data and the model are accessible to the cloud provider in plaintext. Even patient data in the original training dataset could be extracted~\cite{membership-inference-attack}. In addition, the model owner cannot enforce access control, because the cloud provider can share the model to third parties, or execute inference requests from unauthorized users without being detected.  

Serverless computing, the new cloud computing paradigm, provides an opportunity for building model inference services that are cost-effective and quick to scale~\cite{one-step-forward,berkeley-serverless}. Serverless computing allows  
the model owner to focus on the application logic and leave the infrastructure-level tasks, such as resource management
and auto-scaling, to the platform provider. It offers a fine-grained pricing model in which the user is only charged for
the actual resources consumed during the application execution.
Previous works have demonstrated the benefits of serverless model inference~\cite{our-work, mark,fsd-inference}, but do not consider the security of the models or request data. 

We design a model inference system that is secure against the untrusted cloud provider, while preserving the
benefits of serverless services on the cloud.  Specifically, we have four design goals. The first is security; that
is, the cloud and unauthorized users cannot see the models and requests in the clear. The second is high performance;
that is, the system achieves low request latency at scale.  The third is cost-effectiveness; that is, the system
reduces the monetary cost for the model owner. The final goal is non-intrusiveness; that is, the system does not involve
any changes to the existing serverless infrastructure. The first three goals remove the barriers to user adoption, while
the fourth makes the system readily deployable by existing cloud providers.

We leverage trusted hardware~\cite{intel-sgx-explained, arm-trustzone, amd-sev} for security and seamless integration in model inference applications.
In particular, we use Intel SGX, which can create trusted execution environments (TEEs) called \textit{enclaves}, within
which both data and computation are protected by the hardware. The enclaves are secure against powerful attackers,
including the cloud provider that controls the entire software stack~\cite{intel-sgx-explained}. However, trusted hardware alone is not sufficient to achieve all four design goals. Three challenges arise from the characteristics of serverless computing, the performance overhead of trusted hardware, and the infrequent and unpredictable workloads. 

\noindent\textbf{Challenge 1: No direct communication channels between users and serverless instances.}
The serverless platform creates and manages ephemeral instances, e.g., in the form of Docker containers, to handle user
requests.  These instances are hidden from the user, that is, there are no direct communication channels between the user
and the instances. However, for security, the user must establish a secure channel with the running enclave via a remote attestation protocol. Thus, the challenge is how to perform remote attestation with the serverless
instance, without having a direct communication channel with the latter.

\noindent\textbf{Challenge 2: Performance overhead of trusted hardware.}
Trusted hardware introduces two sources of performance overhead. First, setting up trusted execution environments is costly~\cite{sgx2-benchmark}. For example, Intel SGX needs to allocate and initialize the reserved memory regions for the enclave codes and data~\cite{intel-sgx-explained}. 
Second, the overhead associated with mechanisms to ensure security when codes inside TEE interact with non-TEE software, such as encryption and remote attestation, is significant. 
Therefore, the challenge is how to achieve low request latency despite these overheads. 

\noindent\textbf{Challenge 3: Infrequent and unpredictable workloads.} 
Online services that target individual users often experience infrequent and unpredictable workloads~\cite{serverless-in-the-wild, swayam, mlperf},  depending on the nature of the services.
For example, patients may use a set of ML model services from a hospital at certain times of the day for monitoring, or whenever they encounter discomfort. Such workloads reduce the benefits of reusing warm instances, since the serverless platform needs to launch separate instances to handle requests for different models.
As a consequence, each request incurs the cost of cold starting and long initialization time, especially when using TEEs,
which translates to high monetary costs. Therefore, the challenge is to reduce the cost per request under these workloads.

Existing works that integrate TEEs to serverless are not designed for model inference, and they do not achieve all of our
four goals. In particular, \cite{s-faas, securev8,clemmys, reusable-enclave} allow users to attest the serverless instances before
execution. However, they only protect the privacy of user requests and do not consider access control and privacy of models that should be encrypted and loaded on demand.
Furthermore, they do not consider the performance overhead of running model inference inside TEEs, thus they suffer from poor performance.   
Finally, they require extensive changes to existing serverless platforms~\cite{securev8, clemmys, reusable-enclave}, or SGX~\cite{PIE}, which fail to meet our fourth goal. 

We present a novel secure serverless model inference system, called \name{}, that achieves the four design goals.
The system consists of three key features.
The first feature, called \ksname{}, enables secure channels between authorized users and serverless
instances, therefore achieving the first goal. The next feature is an enclave runtime, called \rtname{}, which reuses
the enclave memory for serving multiple requests for the same model, and enables concurrent request execution.  \rtname{}
helps reduce the number of cold starts, which improves the model inference
performance and therefore meets the second goal. The final feature, called \fpname{}, allows multiple models to reuse the same instance, which improves resource sharing under infrequent requests, while retaining the ability to scale when the request rate increases.
By sharing resources among multiple models, \fpname{} addresses the third challenge, and enables \name{} to meet the third goal.
\ksname, \rtname, and \fpname{} involve no changes to the existing serverless software stack, except for the SGX support, which means \name{} meets the fourth design goal.

In summary, we make the following contributions.
\begin{itemize}[topsep=2mm,leftmargin=15pt]
\item We present a novel serverless model inference system, \name{}, that achieves security, low latency, and low cost.
It incurs no changes to existing serverless infrastructure.  
\item We design a trust establishment module, called \ksname{}, bridging the users and serverless instances to build secure channels for remote attestation and access control. 
\item We present a novel enclave runtime that addresses the performance overhead of trusted hardware. The runtime
supports concurrent request execution, and it amortizes the initialization cost over multiple requests.
\item We present a model management component, called \fpname, that reduces the monetary cost of running model inference, by sharing resources across multiple models.
\item We implement \name{} on top of Apache OpenWhisk, a state-of-the-art open source serverless platform. We conduct
extensive experiments with three representative ML models, two inference frameworks, and two hardware versions of SGX, demonstrating that \name{} achieves high
performance and low cost for model inference workloads.
It is open-sourced at \href{https://github.com/nusdbsystem/SeSeMI}{https://github.com/nusdbsystem/SeSeMI}.
\end{itemize}

The remainder of the paper is structured as follows. Section~\ref{sec:background} introduces the background. 
Section~\ref{sec:overview} presents the overview of
\name{}.  Section~\ref{sec:design} describes the design in detail, followed by the implementation in
Section~\ref{sec:implementations}. Section~\ref{sec:evaluation} presents the evaluation results.
Section~\ref{sec:related-work} discusses the related work, before Section~\ref{sec:conclusion} concludes.

\section{Background}\label{sec:background}
\subsection{Trusted Execution Environment}
\label{subsec:trusted-hardware}

Trusted hardware reduces the application's trusted computing base (TCB), by enabling trusted execution environments
(TEEs) that are secure against privileged software attackers. Only the code and data inside a TEE need to be trusted.
Examples of trusted hardware include Intel Software Guard Extensions (SGX)~\cite{intel-sgx-explained}, Intel TDX~\cite{intel-tdx}, AMD Secure Encrypted Virtualization (SEV)~\cite{amd-sev}, ARM TrustZone~\cite{arm-trustzone}, and RISC-V Keystone~\cite{keystone}.
SGX is the most popular, and its TCB consists of only the user code and data~\cite{intel-sgx-explained}, which is significantly smaller than alternative solutions that protect virtual machines~\cite{intel-tdx, amd-sev}. It has been widely used to provide security for data management in untrusted environments~\cite{vc3, opaque, azure-sql-ae, operon, veridb, tgcb, veritxn, data-station}.
Many cloud providers, such as Microsoft Azure,
Alibaba, and IBM, offer SGX-enabled servers. 
SGX provides two key features for securing
applications. One is the trusted execution environment, called {\em enclave}, the other is \textit{remote attestation}, which allows the enclave to prove to a third party that it is launched by a valid SGX platform with up-to-date configurations and is running the correct code and data via a hash value, MRENCLAVE, that can be independently derived. 
Remote attestation can establish a secure channel with an enclave~\cite{ratls}. Intel affirms its support for SGX on its Xeon processors for enterprise and cloud use~\cite{intel-continue-support-sgx}. The enclave page cache (EPC) available for setting up TEE is up to 512GB per CPU, allowing hosting of multiple ML models per server.

Developing an enclave application requires partitioning the application logic into a trusted part, which runs inside the enclave, and an untrusted part that interacts with the OS. 
The enclave developer defines a set of functions, \textsc{ecalls} and \textsc{ocalls}, that enable interaction between the trusted and untrusted code. There are frameworks for running unmodified
applications inside the enclave, by linking the application with a library OS or shim libraries inside the
enclave~\cite{Scone, graphene-sgx, sgx-port-code}. However, these frameworks increase the enclave TCB, which leads to a larger attack surface~\cite{asyncshock, coin-attack}.
Intel SGX enclaves support multi-threading. Although threads cannot be created inside an enclave, multiple threads launched outside of an enclave can invoke enclave functions concurrently. SGX uses a thread control structure (TCS) for each thread to maintain its execution context inside the enclave~\cite{intel-sgx-explained}.

\subsection{Serverless Model Inference}
\label{subsec:serverless-computing}

Serverless computing is an emerging cloud-native service paradigm. Initially offered on the public cloud for running stateless functions, it has recently been extended to support more complex and general workloads~\cite{aft, netherite}, including data analytics~\cite{starling-serverless, lambada}, workflow execution~\cite{cloudburst, boki}, and machine learning~\cite{mbs, infless, mark, fsd-inference}.
It has two distinguished features: ease of management, and
fine-grained pricing. First, the user only defines the application logic and the trigger that invokes the
application~\cite{serverless-in-the-wild}. The cloud provider then takes care of
managing the computing instances, for example, Docker containers, and scheduling the requests to be executed by the
instances.  Second, the user is charged based on the actual resources consumed, for example, by how many milliseconds
the application takes.

\begin{figure}[t]
    \centering
    \includegraphics[width=0.48\textwidth]{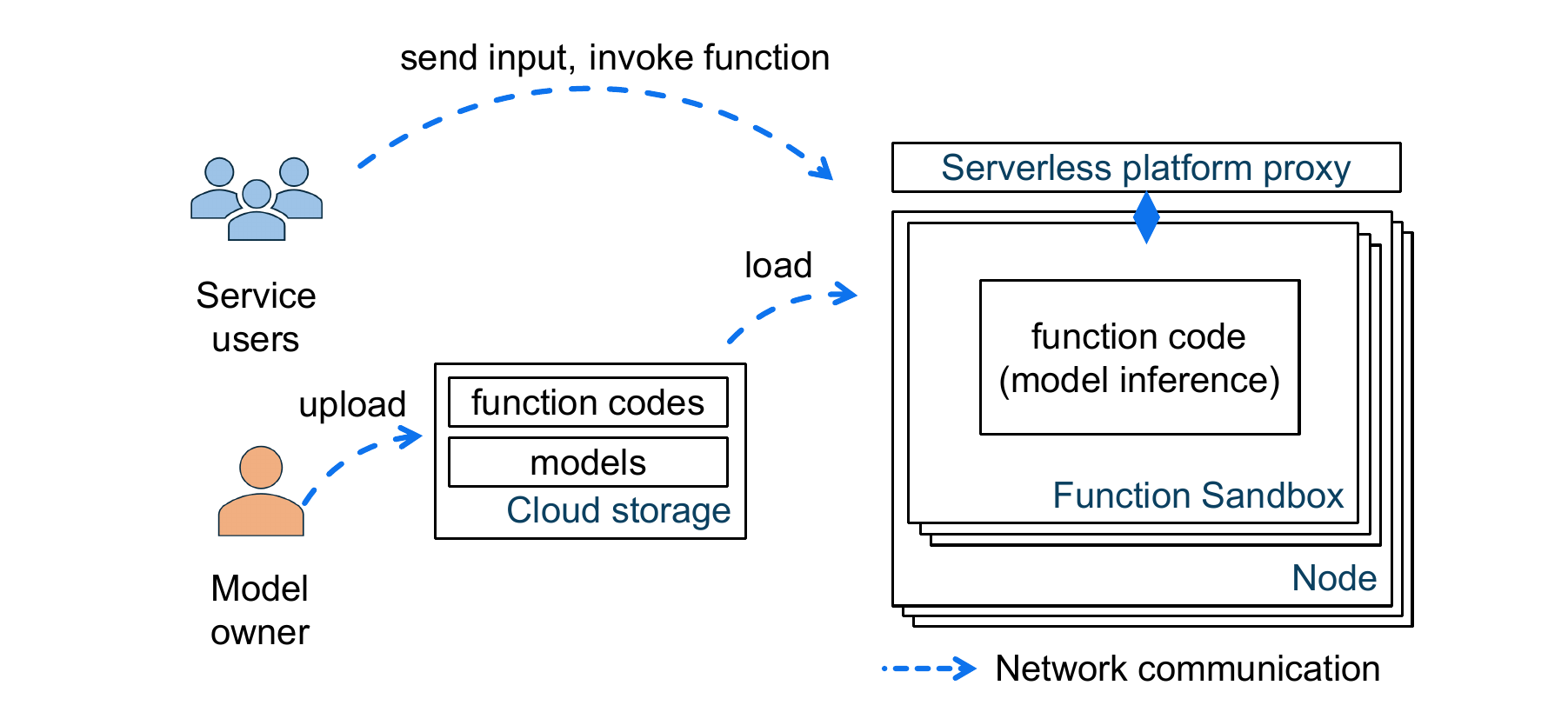}
    \caption{Example of serverless model inference.}
    \label{fig:serverless-sample}
\end{figure}

Figure~\ref{fig:serverless-sample} shows an example of running model inference in a serverless function.  
The model owner first uploads the model to a cloud storage system, then deploys a function with a specific inference runtime.  Most serverless
platforms support two mechanisms for function deployment. The first is the function source code in a high-level language
such as Python, Go, and Javascript, together with the dependency packages. The second is a complete container image,
which gives the model owner more control over the dependencies and the programming languages. Once deployed, the user is given an HTTP endpoint associated with the function. When a request is received at the endpoint, the platform pulls the function code or
container image from the cloud storage, and provisions a sandbox instance to execute the function, passing it the
request data. Before the actual model inference, the function downloads the models and initializes the runtime. These
two steps add significant overhead to the overall latency~\cite{our-work}.

\section{Overview}
\label{sec:overview}

In this section, we provide an overview of \name{} for secure serverless model inference on the cloud. We first describe the system model and threat model, introduce our design goals, and then present the architecture and its overall workflow.

\noindent
\textbf{System model.} There are three main entities: the \textit{model owner}, the \textit{model user}, and
the \textit{cloud}.
The model owner trains a machine learning (ML) model using private data and deploys the model to the cloud. The model
user makes inference requests to the deployed model. For example, the model user can be a patient who uses a
model developed by a hospital to predict her diagnosis given the medical records. The cloud manages the serverless
infrastructure, providing both storage and computing resources to the model owner and user. We assume that the cloud
uses SGX-enabled hardware. 

\noindent
\textbf{Threat model.}
The cloud is not trusted to protect the privacy and integrity of the model and user requests.  In particular, it can
be compromised, either by malicious insiders or by exploiting software vulnerabilities.
Once compromised, the entire software stack, including the operating system and hypervisor, is under the adversary's control, except for the hardware enclaves, which we assume SGX correctly implements the isolation and attestation primitives.  
The adversary can load arbitrary enclaves and invoke arbitrary sequences of enclave functions. 
We do not consider side-channel attacks, physical attacks, or denial-of-service attacks against SGX, which is a common assumption in the literature~\cite{seccask, slalom, operon, tgcb, SecureTF}.
We assume that the code running in the enclave contains no vulnerabilities. 
This can be achieved using static analysis, since the code only loads the model and executes inference. 

\noindent\textbf{Design goals.} We build a serverless model inference system, called \name{}, that
achieves the following four goals.
\begin{itemize}[leftmargin=*]
    \item \textit{Security.} \name{} ensures that the adversary cannot learn the model parameters or the content of
user requests. It prevents inference executions from unauthorized model users.
    \item \textit{High performance.} \name{} ensures low latency for user requests, and can scale to handle high
workloads. In particular, the system incurs a small number of cold starts.
    \item \textit{Cost effectiveness.} \name{} incurs low monetary costs for the model owner. In particular, it aims to
reduce the memory consumption per request.
    \item \textit{Non-intrusiveness.} \name{} does not require any changes to the current serverless infrastructure.
\end{itemize}

\begin{figure}[t]
    \centering
    \includegraphics[width=0.48\textwidth]{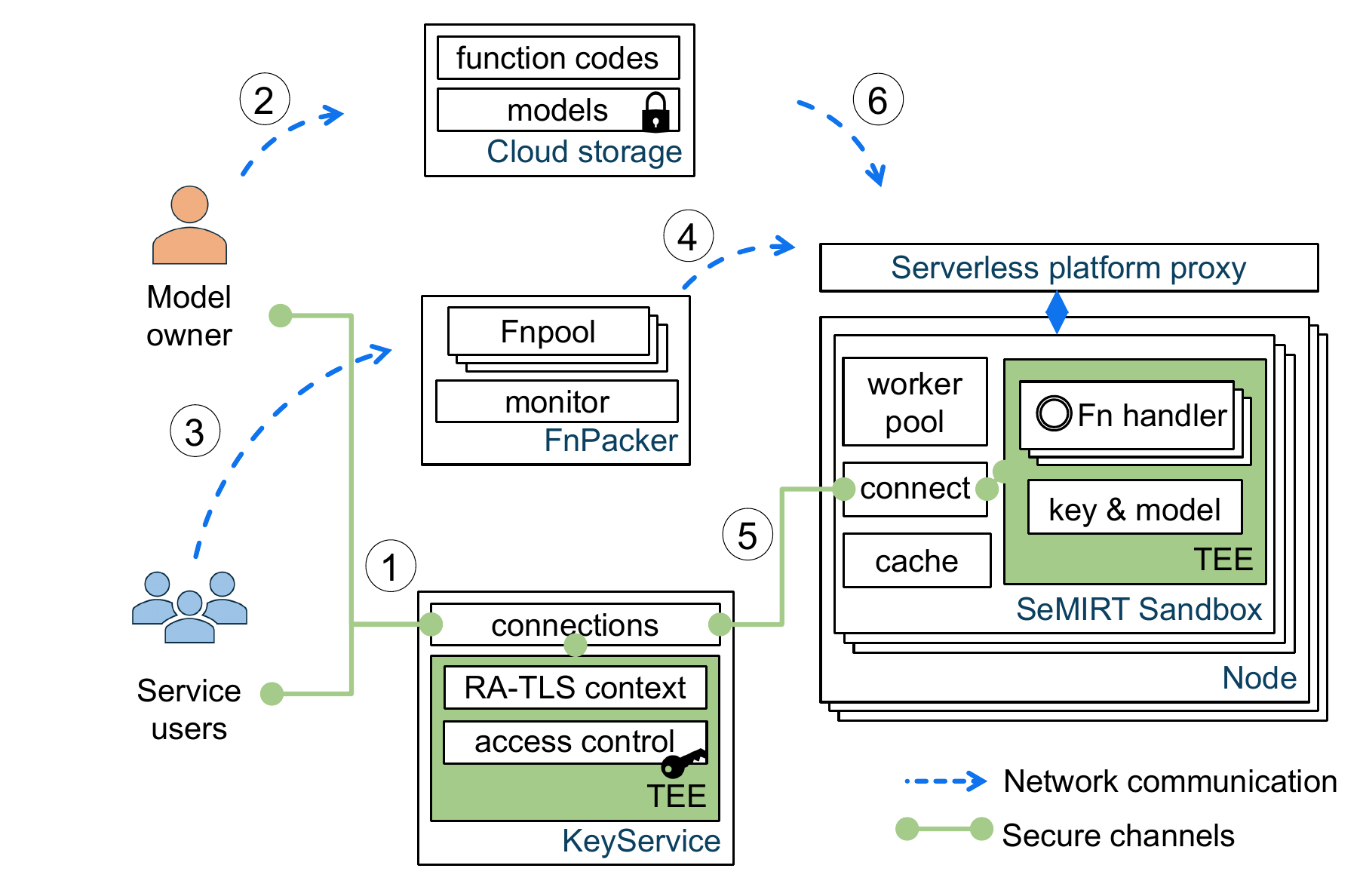}
    \caption{\name{} overview.}
    \label{fig:\name{}}
\end{figure}

\noindent\textbf{Architecture.} Figure~\ref{fig:\name{}} shows a high-level architecture of \name. It consists of three
new components on top of an existing serverless platform.
\begin{itemize}[leftmargin=*]
    \item \ksname{}: it serves as a bridge between the user and the serverless instances. The model owner and user
attest the \ksname{}, which in turn attests the instances and provisions the decryption keys for the
model and the requests to the correct instances. Both the owner and user can specify access control on their decryption keys. 
    \item \rtname{}: it is a tailored enclave runtime that can be deployed as a container image. It supports efficient concurrent inferences inside the enclave. 
    \item \fpname{}: it keeps track of the endpoints that can serve multiple models, and schedules requests to these endpoints based on their workloads.
\end{itemize}

\noindent\textbf{Workflow.}
Figure~\ref{fig:\name{}} depicts how the model owner and user interact with \name{} in three stages: key setup, service deployment, and request serving.  
In the key setup stage (step 1), both the owner and the user attest \ksname, then register their identities and long-term
keys with it. Next, the owner generates a \textit{model key} for encrypting the model, and the user generates a
\textit{request key} for encrypting the requests. These keys are sent to \ksname{} via secure channels. The owner
and user can specify
an access control policy in \ksname, allowing only enclaves with a certain identity $E_S$ to be able to access the
decryption keys. 
The enclave identity --- the MRENCLAVE value --- is based only on the codes for loading and executing the model, not on the model content. Given the codes, the model owner and users can derive $E_S$ independently.
In the service deployment stage (step 2), the owner encrypts the model and uploads it to the cloud storage. 
After that, she deploys the function codes that use \rtname{} as the runtime for model inference.
Finally, the owner deploys the \fpname{} service.
In the request serving stage (steps 3-6), the user encrypts her request and sends it 
to \fpname{}, which forwards it to the serverless proxy. The platform then creates a new instance for handling the request, or forwards it to an active instance. 
For a new instance, \rtname{} launches an enclave and loads the encrypted model and encrypted request into the enclave. Next, it retrieves the decryption keys from \ksname{} via a mutual remote attestation protocol (step 6).
Thereafter, the enclave decrypts the model and request, and executes the inference. The result is
encrypted with the same request key and sent back to the user. 

During the entire process, the user requests and the model are always encrypted except for when they are inside the TEEs. The codes running inside \rtname{} and \ksname{} enforce access control policies stored in \ksname{}. As a result, model inference is performed only if the execution is authorized by both the model owner and the user. 

\section{\name{} Design}\label{sec:design}

This section presents the \name{} design. Specifically, we elaborate on the design of the three components \ksname{}, \rtname{}, and \fpname{} in Sections~\ref{subsec:\ksname{}}, \ref{subsec:\rtname{}}, and \ref{subsec:\fpname{}}. 

\subsection{KeyService} \label{subsec:\ksname{}}
To meet the security goal, it is necessary to encrypt both the models and the requests so that they can only be decrypted
inside a correct enclave. Furthermore, the decryption keys are only given to enclaves processing authorized model users' requests. The
characteristics of serverless computing make it difficult to realize this goal. In particular, there are no direct
communication channels to the serverless instances (only the serverless proxy knows where the instances are), thus the
owner and user cannot perform remote attestation with the enclaves.
We address this problem by introducing a new component \ksname{} that serves as a bridge to the enclaves. \ksname{} is
an always-on service that consists of an enclave with two main functionalities. First, it manages the decryption keys
and enforces access control policies. Second, it provisions the keys to the correct enclaves. 

\noindent\textbf{Design.} \ksname{} enclave stores four sets of data.  
\begin{itemize}[leftmargin=*]
    \item $KS_I$: the set of $\langle id, K_{id} \rangle$ tuples, where $id$ and $K_{id}$ denote the owner or user
identity, and their corresponding long-term key.  
    \item $KS_M$: the set of $\langle M_{oid}, K_{M} \rangle$ tuples, where $M_{oid}$ and $K_M$ are the model identity
owned by $oid$, and the model decryption key, respectively. 
    \item $KS_R$: the set of $\langle M_{oid} \Vert E_S \Vert uid, K_{R} \rangle$ tuples, specifying that the request key
$K_R$ from user $uid$ can only be given to the enclave that has identity $E_S$ and running model $M_{oid}$.  
    \item $AC_M$: the set of $\langle M_{oid} \Vert E_S \Vert uid \rangle$ records, specifying that the decryption key
for $M_{oid}$ can only be given to the enclave that has identity $E_S$ and is handling user $uid$'s requests.  
\end{itemize}

\begin{algorithm}[t]
\SetKwFunction{algo}{algo}
\SetKwFunction{userreg}{\textsc{user\_registration}}
\SetKwFunction{addobj}{\textsc{add\_model\_key}}
\SetKwFunction{grantac}{\textsc{grant\_access}}
\SetKwFunction{addinput}{\textsc{add\_req\_key}}
\SetKwFunction{handleworker}{\textsc{key\_provisioning}}
\SetKwProg{myalg}{Algorithm}{}{}
\SetKwProg{myproc}{Procedure}{}{}
\DontPrintSemicolon
\small
{
$KS_{I}$: storage for owner and user's identity keys \\
$KS_{M}$: storage for model decryption keys \\
$KS_{R}$: storage for access control of request keys \\
$AC_{M}$: storage for model access control \\
\myproc{\userreg{$K_{id}$}}{
  $id \leftarrow SHA256(K_{id})$ \\
  $KS_{I} \leftarrow KS_{I} ~\bigcup~ \langle id, K_{id} \rangle$ \\
  \KwRet $id$\;
}
\myproc{\addobj{$oid$, $[M_{oid} \Vert K_{M}]_{K_{oid}}$}}{
  $K_{oid} \leftarrow$ get the identity key of $oid$ in $KS_{I}$ \\
  $M_{oid}$, $K_{M}$ $\leftarrow$ decrypt $[M_{oid} \Vert K_{M}]_{K_{oid}}$ using $K_{oid}$ \\
  $KS_{M} \leftarrow KS_{M} ~\bigcup~ \langle M_{oid}, K_{M} \rangle$
}
\myproc{\grantac{$oid$, $[M_{oid} \Vert E_S \Vert uid]_{K_{oid}}$}}{
  $K_{oid} \leftarrow$ get the identity key of $oid$ in $KS_{I}$ \\
  $M_{oid}$, $E_S$, $uid$ $\leftarrow$ decrypt $[M_{oid} \Vert E_S \Vert uid]_{K_{oid}}$ \\
    $AC_{M} \leftarrow AC_{M} ~\bigcup~ \langle M_{oid} \Vert E_S \Vert uid \rangle$
}
\myproc{\addinput{$uid$, $[M_{oid} \Vert E_S \Vert K_{R}]_{K_{uid}}$}}{
  $K_{uid} \leftarrow$ get the identity key of $uid$ in $KS_{I}$ \\
  $M_{oid}$, $E_S$, $K_R$ $\leftarrow$ decrypt $[M_{oid} \Vert E_S \Vert K_{R}]_{K_{uid}}$ \\
    $KS_{R} \leftarrow KS_{R} ~\bigcup~ \langle M_{oid} \Vert E_S \Vert uid, K_{R} \rangle$
}
\myproc{\handleworker{$uid$, $M_{oid}$, $RAReport$}}{
  $E_S \leftarrow$ get enclave identity from $RAReport$ \\
  \If{$\langle M_{oid} \Vert E_S \Vert uid \rangle \in AC_{M} \wedge \langle M_{oid} \Vert E_S \Vert uid \rangle \in KS_{R}$}{
    $K_{M} \leftarrow$ get model key of $M_{oid}$ from $KS_{M}$ \\
    $K_{R} \leftarrow$ get request key from $KS_{R}$ \\
    \KwRet $K_{M}$, $K_{R}$
  }
}
}
\caption{KeyService enclave operations}\label{alg:key-service}
\end{algorithm}

The identity of \ksname{} enclave, denoted as $E_K$ is fixed. The owner and user attest \ksname{} before interacting with it. 
Algorithm~\ref{alg:key-service} describes the main operations of \ksname. Inputs to these functions are sent to
\ksname{} via a secure channel. The first four functions update the corresponding sets of data that store the decryption keys and access control policy. 
Specifically, the owner and user need to call the \textsc{user\_registration} function to register their identity keys before using \name{}. \ksname{} provides its remote attestation report to the owner and user. The owner and user check the report's authenticity and integrity and compare the enclave identity in the report with the value $E_K$ they obtained beforehand. After passing the check, they can provide their identity keys to the \ksname{} enclave, which computes the cryptographic hashes of the keys as the owner and user ids, and stores them in $KS_I$.

Next, the model owner can add a model key for each serving model using the \textsc{add\_model\_key} function. Since each model owner may have many ML models, \name{} allows the owner to set different model keys for different ML models. The model id and the corresponding model key $\langle M_{oid}, K_M \rangle$ are stored in $KS_M$. Then, the model owner can authorize the access of model $M_{oid}$ to user $uid$ through the \textsc{grant\_access} function, storing the access control policy $\langle M_{oid} \Vert E_S \Vert uid \rangle$ in $AC_M$, where $E_S$ indicates that only the enclave $E_S$ is valid to retrieve the corresponding keys. Similarly, a user $uid$ adds the request key $K_R$ associated with $\langle M_{oid} \Vert E_S \Vert uid \rangle$ using the \textsc{add\_req\_key} function, storing in $KS_R$. This information prevents the misuse of the user's request key. 

The \textsc{key\_provisioning} function is invoked by an enclave running in a serverless instance to retrieve the decryption keys. The \ksname{} and the requesting enclave first perform mutual
attestation, in which the latter verifies that it is
communicating with $E_K$. \ksname{} checks that the identity of the requesting enclave, $E_S$, is
authorized by the access control policy, by matching it with the $KS_R$ and $AC_M$ and returns the decryption keys to $E_S$ via a secure channel, if authorized.

\subsection{\rtname{}} \label{subsec:\rtname{}}

\begin{figure}[t]
    \centering
    \includegraphics[width=0.48\textwidth]{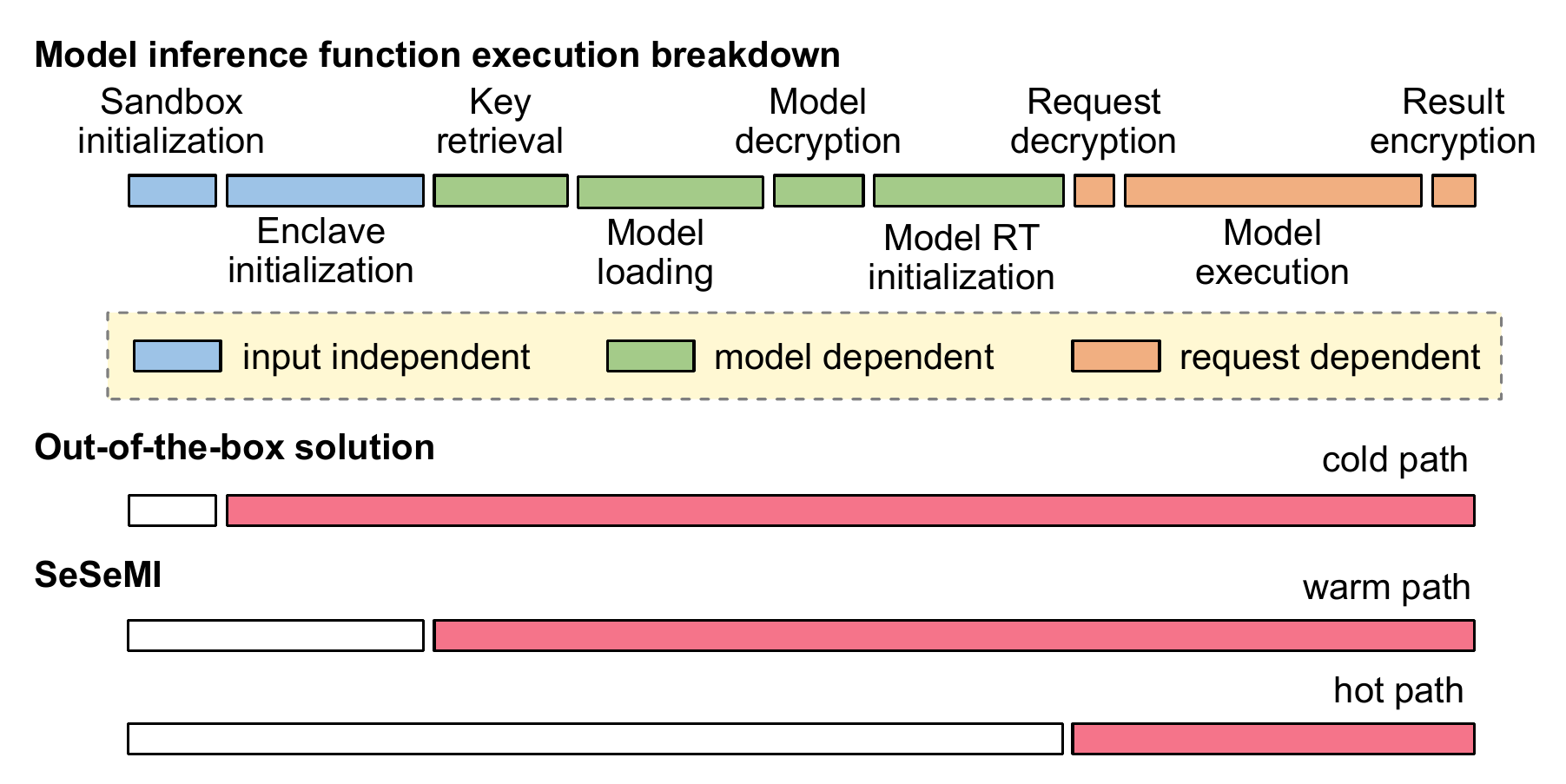}
    \caption{Model serving stages inside a serverless instance.}
    \label{fig:serve-stages}
\end{figure}

When a user invocation is received, the serverless platform can launch an instance to process the model inference request. Figure~\ref{fig:serve-stages} illustrates the steps for a cold-started instance.
The {sandbox
initialization} step includes pulling the Docker image from the cloud storage and starting a new instance. The {enclave
initialization} step initializes a new enclave. 
The key retrieval step performs mutual remote attestation with \ksname{} before getting the corresponding keys. 
The next three steps load the encrypted model from cloud storage and decrypt it inside the enclave. The {model runtime initialization} step starts the model inference runtime. 
The {request decryption} step retrieves the key and decrypts the user request inside the enclave. The last two steps perform model inference, encrypt the results and send them back to the user.
We observe that out of these nine steps, the key retrieval, model loading, model decryption, and model runtime initialization steps are dependent on the serving model, 
while the request decryption, model inference, and result encryption step depend on the individual request data.
In other words, there is an opportunity to amortize the cost of the other steps across multiple requests. 
Existing serverless platform considers a new instance as cold if it has to perform
the sandbox initialization step, and as warm if the instance is already initialized, i.e., it can skip the first step.
As a consequence, the enclave initialization step is executed even for warm instances.  We observe that this cost is high due to the overhead of SGX especially when multiple enclaves are launched at the same time. 

\begin{figure*}[t]
    \centering
    \begin{tabular}{p{0.42\textwidth}p{0.56\textwidth}}
    \hline
    \textbf{ECALL} \\
    \textsc{ec\_model\_inf}($[data]_{K_{R}}$, $uid$, $M_{oid}$, $KeyService$)  & run inference on user $uid$'s input with model $M_{oid}$ in the enclave \\
    \textsc{ec\_get\_output}($buf$) & copy the encrypted output to a buffer in untrusted memory \\
    \textsc{ec\_clear\_exec\_ctx}() & allow untrusted codes to free execution context \\
    \textbf{OCALL} \\
    \textsc{oc\_load\_model}($M_{oid}$) & load an encrypted model from untrusted memory into the enclave \\ 
    \textsc{oc\_free\_loaded}($M_{oid}$) & free the resource hold at the untrusted memory for model loading \\
    \textbf{Inference API} \\
    \textsc{model\_load}($M_{oid}$, $K_{M}$) & load and decrypt the model with id $M_{oid}$ to the enclave \\
    \textsc{runtime\_init}($model$) & initialization of a model runtime \\
    \textsc{model\_exec}($data$, $model$, $model\_rt$) & process the input and calculate the model inference result \\
    \textsc{prepare\_output}($model\_rt$)  & retrieve the output from model runtime and serialize it into a byte buffer\\
    \hline
    \end{tabular}
    \caption{Enclave \textsc{ecall} \& \textsc{ocall} APIs and Inference APIs in \rtname{}.}
    \label{fig:runtime-api}
\end{figure*}

\begin{figure}[t]
    \centering
    \includegraphics[width=0.46\textwidth]{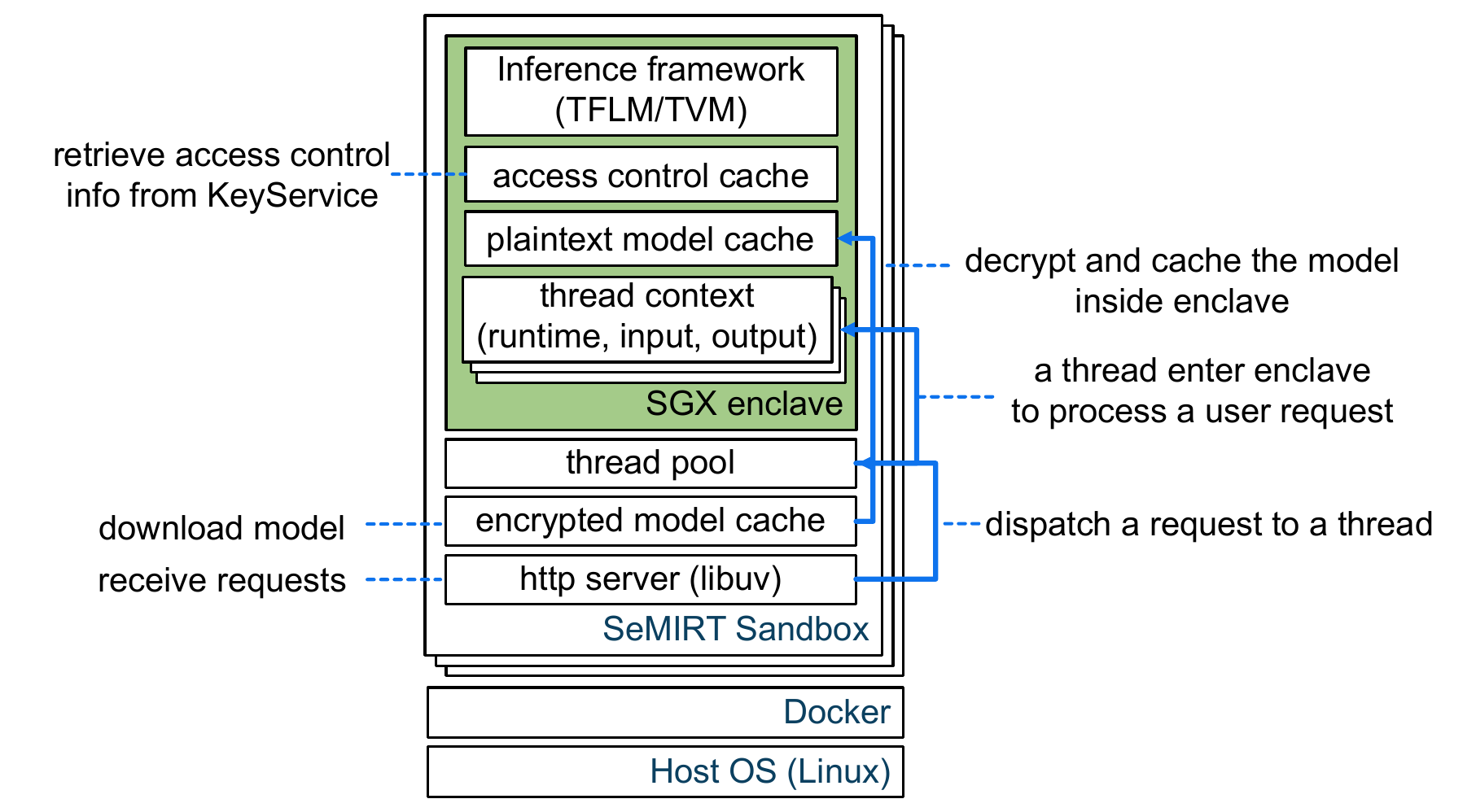}
    \caption{\rtname{} internal.}
    \label{fig:\rtname{}}
\end{figure}

\noindent\textbf{\rtname{}.} We design \rtname{} as a new serverless runtime that reduces the execution time of warm
invocations and the overall memory consumption per enclave. The key techniques include reusing more states from the
previous invocations for warm instances, and executing multiple concurrent requests in the same enclave to reduce memory consumption per request.
 
Out of the nine steps in Figure~\ref{fig:serve-stages}, the sandbox initialization, enclave initialization, and remote attestation steps can be reused for different
models and users. The same loaded model and initialized runtime can be reused for different users. The same model
keys and request keys can also be reused for different requests from the same user. \rtname{} stores some data
from the previous invocations such that they can be reused to speed up the next request. In particular, it stores the previous  
RA report (for SGX1) or RA quote (for SGX2),  
the last $\langle uid, M_{oid} \rangle$ pair, and the corresponding decryption keys, the
decrypted model, and the model runtime. Specifically, \rtname{} handles requests in three ways, as illustrated in
Figure~\ref{fig:serve-stages}.  
\begin{itemize}[leftmargin=*]
    \item \textit{Cold invocation}: a new instance is started from scratch to handle the request, this is the same as existing serverless runtimes. 
    \item \textit{Warm invocation}: the request is handled by a warm instance whose enclave has been initialized, but
the correct model has not been loaded. 
    \item \textit{Hot invocation}: the request is handled by a warm instance that has loaded the same model and the
request key of the same user. 
\end{itemize}

\begin{algorithm}[!t]
\SetKwFunction{algo}{algo}
\SetKwFunction{inf}{\textsc{ec\_model\_inf}}
\SetKwFunction{out}{\textsc{ec\_get\_output}}
\SetKwProg{myalg}{Algorithm}{}{}
\SetKwProg{myproc}{Procedure}{}{}
\DontPrintSemicolon
\small
{
global $Model$ \\
global $KC$: last pair of decryption keys \\
thread\_local $model\_rt$ \\
thread\_local $Output$ \\
\myproc{\inf{$[data]_{K_{R}}$, $uid$, $M_{oid}$}}{
    \If {$M_{oid} \Vert uid \notin KC$} {
        $K_{M}, K_{R} \leftarrow$ retrieve keys from $KeyService$  \\
        $KC \leftarrow \langle M_{oid} \Vert uid, (K_{R}, K_{M}) \rangle$
    }
    \Else {
        $K_{M}, K_{R} \leftarrow$ retrieve keys from $KC$  \\
    }
    \If {$Model$ is not the target model $M_{oid}$} {
        // model switched with lock when not in use \\
        $Model \leftarrow$ \textsc{model\_load}$(M_{oid}, K_{M})$
    }
    \If {$model\_rt$ is not for the target model $M_{oid}$} {
        $model\_rt \leftarrow$ \textsc{runtime\_init}$(Model)$
    }
    $data \leftarrow$ decrypt $[data]_{K_{R}}$ with $K_{R}$ \\
    \textsc{model\_exec}$(data, Model, model\_rt)$ \\
    $result \leftarrow$ \textsc{prepare\_output}$(model\_rt)$ \\
    $Output \leftarrow$ encrypt $result$ with $K_{R}$ \\
    \KwRet\;
}
\myproc{\out{$buffer$}}{
    // buffer is in untrusted memory, Output is in enclave \\
    $buffer \leftarrow Output$
    \KwRet
}
}
\caption{\rtname{} enclave executions}\label{alg:executor}
\end{algorithm}

\noindent\textbf{Concurrent requests handling.}
We observe that the model and the model runtime buffers account for the majority of the enclave memory.  Thus, we design \rtname{} to process multiple requests in the same enclave, reducing the memory consumption per request. 
Figure~\ref{fig:\rtname{}} illustrates how \rtname{} supports concurrent requests. The requests are dispatched to a thread pool outside the enclave. Each thread handles one
request and enters the enclave via a Thread Control Structure (TCS). The TCS defines the private workspace of the
thread, including input and buffers for intermediate results.  One thread loads and decrypts the model, then stores it in
the heap region shared among all threads. 
The enclave maintains a secure channel with \ksname{} after the first remote attestation and caches the keys in the shared heap region.  

Figure~\ref{fig:runtime-api} lists the \rtname{} enclave interfaces and inference APIs, which are compatible with existing model inference frameworks. We support TFLM~\cite{tflm} and Apache TVM~\cite{tvm} for now. Function developers can easily use them to program function logic to load and execute model inference with the selected model inference framework. \rtname{} aims to introduce minimal \textsc{ecalls} and \textsc{ocalls} to reduce the interface between the trusted enclave codes and untrusted application codes. There is a single \textsc{ecall} \textsc{ec\_model\_inf} that contains the model inference logic. A new thread scheduled with a user request invokes \textsc{ec\_model\_inf} to enter the enclave.
Algorithm~\ref{alg:executor} details the enclave execution logic.
Given a request, it first checks if the model key $K_M$ and request key $K_R$ are loaded in the enclave
(we cache only one pair of keys to avoid requests from multiple users running inside the enclave at the same time). 
If not, it fetches the keys from \ksname{} by calling the  \textsc{key\_provisioning} function, then caching them for future requests (lines 6-10). Next, it checks if the currently loaded model is the target model. If not, it calls the \textsc{model\_load} API to load and decrypt the model 
replacing the current one (lines 11-13). It then invokes \textsc{runtime\_init} API to prepare the model runtime if the current one is not for the target model (lines 14-15). Next, it decrypts the request using $K_R$ and invokes \textsc{model\_exec} to perform the inference (lines 16-17). Finally, it encrypts the result using $K_R$ before returning it to the outside of the enclave.
\rtname{} can be extended to support new inference frameworks, by implementing the four inference APIs. \rtname{} will then manage the model serving steps for efficient execution.

\subsection{\fpname}\label{subsec:\fpname{}}

\begin{figure}[t]
    \centering
    \includegraphics[width=0.46\textwidth]{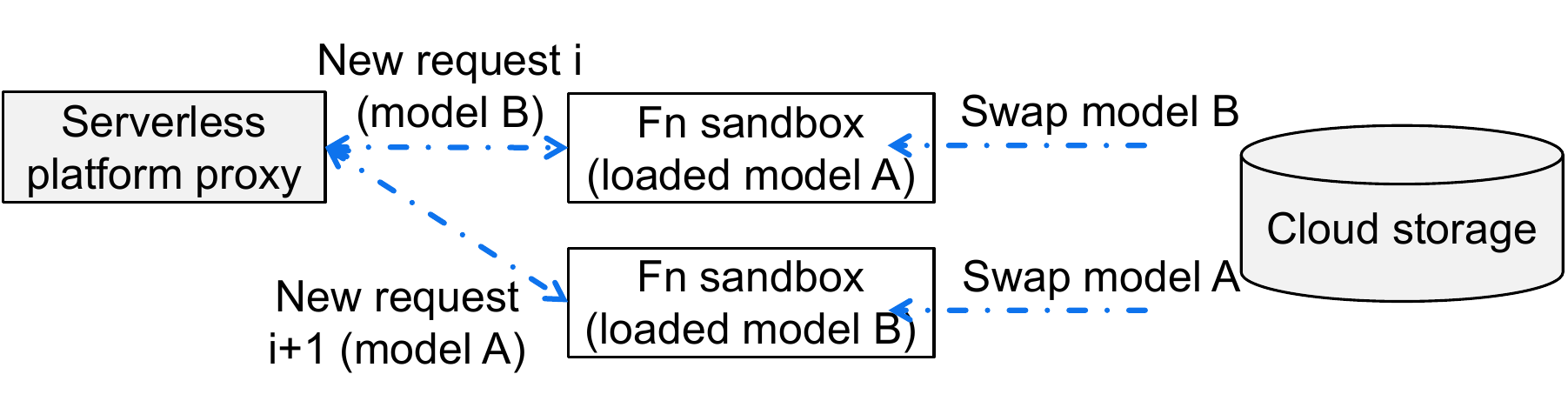}
    \caption{Serverless platform proxy indiscriminately chooses idle sandboxes to serve requests.}
    \label{fig:fnpack-motivation}
\end{figure}

Infrequent and unpredictable workloads are common in serverless computing and online model inference~\cite{serverless-in-the-wild,swayam, mlperf, mbs, spes}. It is challenging to achieve low latency, and thereby low cost, under these workloads, because of the overhead of creating an enclave and initializing the model runtime from scratch.
We observe that a typical model owner needs to manage several similar models for different users. One approach for managing multiple models is to deploy
each model independently, i.e., one endpoint per model. However, this is inefficient when the request rate is low because
a new instance is created from scratch to handle each request, resulting in high latency and monetary costs. The other
approach is to deploy one endpoint that serves all the models (i.e., all-in-one), by switching between different models
inside the same set of sandboxes. Although this approach enables more resource sharing, it becomes inefficient when the
workload is unpredictable. In this case, the cost of switching back and forth the models may outweigh the savings from
reusing the serverless sandboxes, as shown in Figure~\ref{fig:fnpack-motivation}.

We design \fpname{}, a component that groups requests based on the arrival pattern, then routes them 
to suitable serverless endpoints. \fpname{} merges the requests to infrequently accessed
models into a small set of endpoints, and enables those frequently accessed models to use the full resources (without
sharing) for better performance.  
To use \fpname{}, the model owner specifies a \textit{Fnpool} structure that contains a set of models and the memory budget
for an instance. \fpname{} automatically deploys the models to different endpoints, and schedules user requests to
them. \fpname{} monitors the requests and the executions at different endpoints. Its schedule is based on two metrics.
One is the model execution statistics, including the number of concurrent requests pending response on each model, the
last invocation time, and the latency of different types of execution (i.e., cold, warm, and hot). The other is the
state of each endpoint: either idle or busy. 

For a request to a model with pending responses, \fpname{} forwards it to the endpoint for that model, marking this
endpoint as exclusive to this model. For a request to a model without pending responses, it is forwarded to the first
endpoint that is not busy serving another model. An endpoint is
considered not busy when it meets either of the following conditions. First, there is no pending response on the endpoint, and
it is not marked exclusive for another model. Second, it is marked as exclusive, but a large interval has passed since the
last request was sent to it. This simple but effective scheduling strategy helps reduce the number of cold starts by packing requests
to infrequently accessed models to the same endpoint. It does not affect models with high request rates, as they are
marked as exclusive and do not incur the overheads of switching models. 

\subsection{Security Analysis}\label{subsec:security-analysis}

\noindent{\bf Model and request confidentiality.}
The model and request encryption keys are generated by the model owner and user independently. The enclave identities of \ksname{} and \rtname{} are also derived independently by the model owner and users, given only the codes for initializing and running model inference. The encryption keys are provisioned to the \ksname{} via the secure channel established only after a successful remote attestation. The access control list is also provisioned via this secure channel. The encryption keys can only be retrieved by \rtname{} enclaves that are allowed by the access control list. Thus, the keys are protected from unauthorized users and network adversaries. We note that for added protection and performance, multiple \ksname{} can be deployed to isolate keys from different users, which require users to specify the address of the corresponding \ksname{} in their requests.
Considering an attacker who tries to tamper with $KS_R$ and $AC_M$, for example, to add himself to the lists of authorized users. This attack is impossible because the functions that modify $AC_M$ and $KS_R$ ($\textsc{grant\_access}$ and $\textsc{add\_req\_key}$ in Algorithm~\ref{alg:key-service}) check that the updates are authorized, i.e., signed with the long-term key of the model owner $oid$ and the user $uid$, respectively. As a result, \ksname{} prevents unauthorized users from using the model. 

\noindent{\bf \rtname{} security.}
\rtname{} only provides three \textsc{ecalls} into the enclave, and two \textsc{ocalls} from the enclave to the untrusted environment, apart from those in Intel SGX SDK and the remote attestation library. This small interface helps minimize the attack surface. The attacker can only influence the enclave execution through the encrypted and authenticated requests and models. \rtname{} checks the integrity and freshness of the data, then decrypts them with the appropriate keys. As a result, the attacker cannot tamper with the data or perform rollback attacks.
\rtname{} prevents executing requests from unauthorized users, because when the enclave can reuse decryption keys from the previous requests, it means that the enclave itself was
authorized to access the model and the user requests.
\rtname{} avoid potential problems (for example, side-channel leakage) from sharing an enclave across different models and users by running only one model and serving requests from one user per enclave. During model inference, the input, intermediate data, and the result are kept in the thread local storage region. 
\rtname{} further provides the option of disabling concurrent execution, thus ensuring request isolation at the expense of performance.  We explain how to enable sequential processing in Section~\ref{sec:implementations},  and discuss its performance impact in Section~\ref{subsec:micro-bench}.

\noindent{\bf \fpname{}.} We note that \fpname{} does not affect security, because it only routes encrypted requests based on model ID, which is not sensitive. The result sent back to user through it is also encrypted. 

\noindent{\bf Discussion.}
We consider side-channel attacks~\cite{side-channel-attack-survey, foreshadow} out of scope --- a common assumption in TEE-based data management systems~\cite{veridb, seccask, azure-sql-ae, operon, tgcb, ironsafe, ccf}. Nevertheless, we note that Intel routinely releases microcode or hardware updates 
(e.g., Intel-SA-00115, Intel-SA-00161, Intel-SA-00289) to fix side-channel attacks (e.g., CVE2018-3615, CVE-2019-11157)~\cite{foreshadow, plundervolt} . For neural network models, we also note that many computations are data oblivious, and operations that are not oblivious can be rewritten on a case-by-case during the model development phase~\cite{slalom, oblivious-ml}. Consequently, the side channels can be mitigated in most applications running on \rtname{}.

Model-centric attacks, such as data reconstruction attacks~\cite{FredriksonJR15}
and model extraction attacks~\cite{ml-attack-model-extract-1} are also out of scope. However, we note that our system restricts the model access to only authorized users, thus limiting the entities that launch the attacks. Furthermore, we can mitigate model-centric attacks by rounding confidence scores of the prediction output or injecting random noise that satisfies differential privacy into the training phase. 
\section{Implementations}
\label{sec:implementations}
We implement \ksname{} and \rtname{} with 4200 LoC in C/C++, and \fpname{} with 900 LoC in  Go. 
We implement the enclave logic with Intel SGX SDK 2.14. 
We implement remote attestation between \ksname{} and client, and mutual attestation between \ksname{} and \rtname{} based on the RA-TLS implementation~\cite{ratls-code}. 
We build the system on top of Apache OpenWhisk~\cite{openwhisk},  by implementing its action interface for \rtname{}.
OpenWhisk is one of the earliest and most successful serverless platforms, an open-sourced version of the platform backing IBM Cloud function.
Other serverless platforms can be supported by extending their APIs.
We use AES-GCM for model and request encryption. 

\noindent\textbf{\ksname{}}.
The \ksname{} implementation adds only 860 LoC of C/C++ to the TCB, besides the dependencies. It supports multiple connections, and each connection is handled by a thread, which corresponds to a TCS inside the enclave.

\noindent\textbf{\rtname{}.}
We support two popular model inference frameworks, Apache TVM@v0.9~\cite{tvm} and TFLM@d1a2e3d~\cite{tflm}. We patch and extend them to execute concurrent requests inside the enclave. \rtname{} adds 780 LoC to the TCB, on top of the inference frameworks. 
We use libuv~\footnote{\url{https://libuv.org/}} and llhttp~\footnote{\url{https://llhttp.org/}} to implement an asynchronous server conforming to the OpenWhisk specified action interface. They are not part of the TCB. We use the libuv thread pool and assign each request to one thread which is bound to a TCS inside \rtname{} enclave. 
We build \rtname{} as a container image such that it is readily deployable on existing serverless platforms.
The user can configure the concurrency level by setting the number of threads (or TCS) in the enclave configuration file. For example, to enforce strong isolation between requests, which mitigates side-channel attacks, TCS can be set to 1. Furthermore, \rtname{} can be configured to fix the model and disable the key cache, which helps avoid potential side channels arising from sharing enclave resources. The settings are part of the enclave codes.

\noindent\textbf{Clients.}
Model owners and users use a client with RA-TLS to establish secure channels to register, transfer keys, and update access control policies with \ksname{}.
For user requests, we extend the Go client of OpenWhisk, making it communicate with \fpname{} directly. \fpname{} then forwards the requests to OpenWhisk's endpoints based on the scheduling decisions. 

\section{Evaluation}
\label{sec:evaluation}

In this section, we discuss the performance of \name{}. We first evaluate the \rtname{} with micro-benchmarks to demonstrate its benefits on warm and hot invocations. Then, we deploy \name{} with Apache OpenWhisk and evaluate it using realistic workloads~\cite{mark, our-work, BATCH-serve, mlperf}. We conduct experiments in the single-node setting to understand the impact of different components in \name, and experiments in the multi-node setting to understand its scalability. 

\begin{figure*}[t]
  \centering
  \begin{minipage}[b]{0.33\textwidth}
    \includegraphics[width=\textwidth]{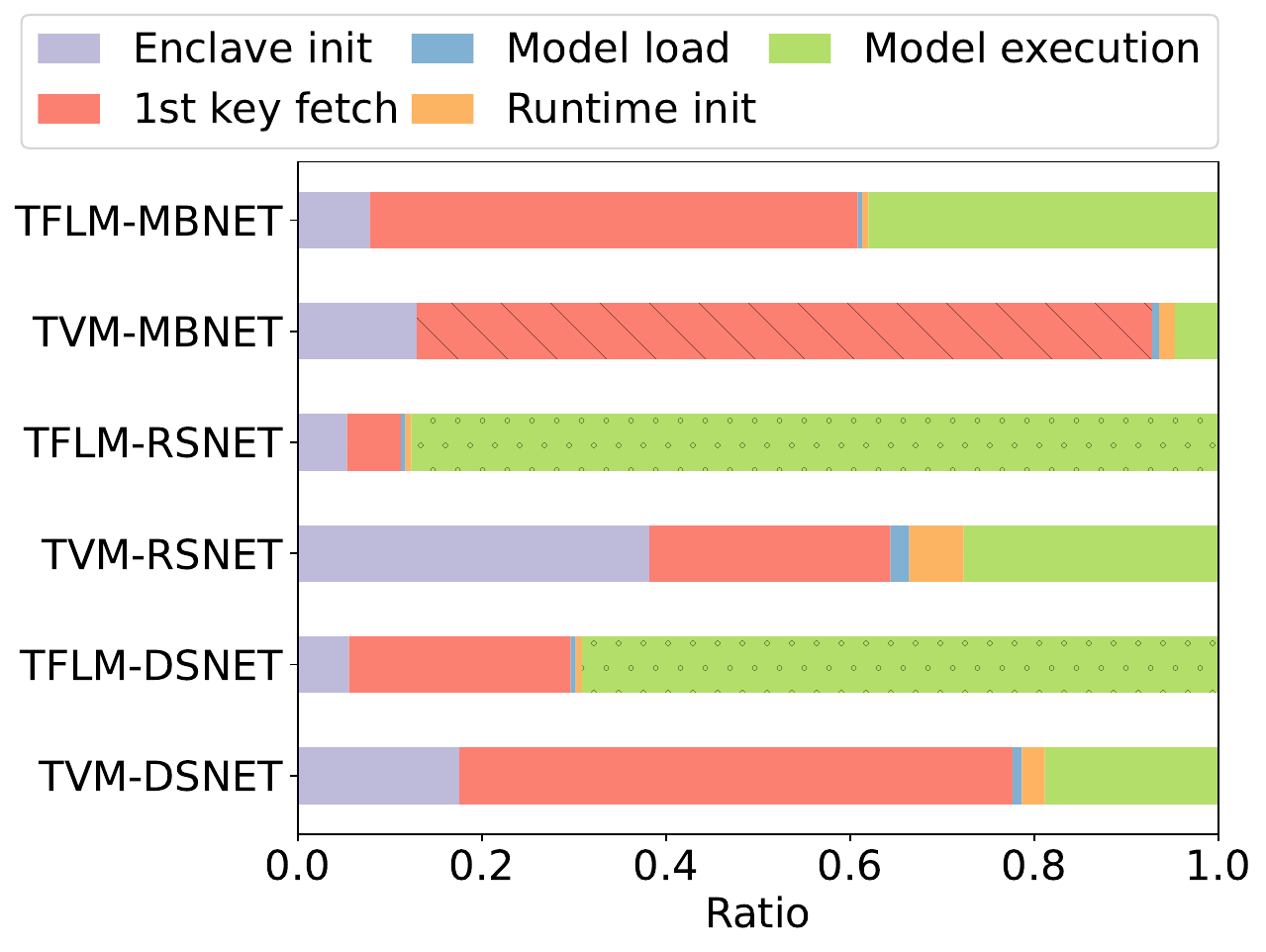}
    \caption{Latency ratio of serving stages.} 
    \label{fig:ratioperf}
  \end{minipage}
  \begin{minipage}[b]{0.64\textwidth}
    \includegraphics[width=\textwidth]{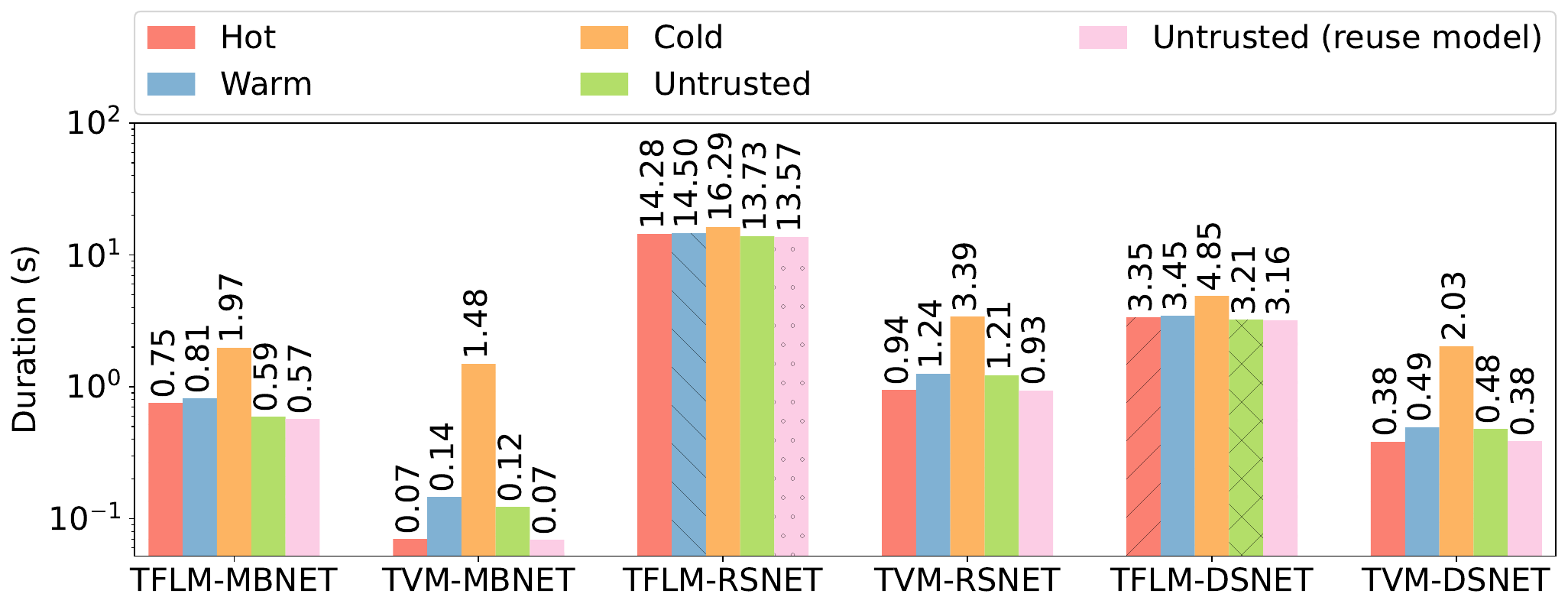}
    \caption{Execution time under different invocations.}
    \label{fig:sumperf}
  \end{minipage}
\end{figure*}

\noindent\textbf{Experimental Setup.}
We use an 11-node cluster for our experiments and each node has Intel Xeon Gold 5317 CPU @ 3.00GHz with 12 physical cores, 128 GB of RAM, and 960 GB SSD. The nodes in the cluster are connected via 10Gbps Ethernet. Intel SGX2 hardware is enabled and configured with the maximum supported EPC size of 64 GB. These nodes run Intel SGX aesmd services based on Linux-sgx 2.14 for launching enclaves and for remote attestation. They run Ubuntu 20.04 with Linux kernel 5.13.0-41.

We set up Apache OpenWhisk with Kubernetes on the cluster and make 8 nodes available to schedule function instances. One node is reserved for the Kubernetes control plane. One node is reserved to run the \ksname{} and \fpname{}, and the remaining node is for issuing requests.
We provision the keys and access control list to the \ksname{} enclave and register the function to \fpname{} following the  workflow described in Section~\ref{sec:overview}.
We also run a set of experiments on a cluster of SGX1 nodes to evaluate the performance of \name{} when the enclave page cache (EPC) available for setting up TEE is limited. This is relevant as the EPC of SGX2 can also be under pressure when multiple large ML models are hosted on a machine. In this setting, the nodes are equipped with Intel Xeon W-1290P CPU @ 3.70GHz and configured EPC size of 128 MB.
A network file system is set up in the cluster to emulate cloud storage. Unless otherwise stated, the results presented in this section are for SGX2 nodes.

\noindent\textbf{Baselines.}
We compare \name{} against two baselines.
\begin{itemize}[leftmargin=*]
    \item \textit{Native}: this baseline follows the design of existing serverless sandbox runtimes such that the warm invocations only reuse the initialized sandbox (see Figure~\ref{fig:serve-stages}), and each sandbox launches a new enclave to handle the invocation. It uses a Docker image that implements a server and runs compiled code against requests.
    \item \textit{Iso-reuse}: this baseline implements the design in~\cite{s-faas, clemmys}. We note that the source code of Clemmy~\cite{clemmys} is not available, while the implementation of S-FaaS~\cite{s-faas} only targets Javascript functions. As a consequence, we implement their performance optimization according to their descriptions in the paper. Specifically,  in this baseline,  warm invocations reuse the initialized enclave and decryption keys inside the enclave. However, the models and inference runtime are loaded and initialized from scratch. 
\end{itemize}

\subsection{\rtname{} Micro-benchmarks} \label{subsec:micro-bench}
We evaluate \rtname{} with three popular models of different sizes: MobileNet v1 (MBNET), ResNet101 v2 (RSNET), and DenseNet121 (DSNET) with the two inference frameworks, namely TFLM and TVM. 
Table~\ref{tab:models} lists the size of the model and the runtime allocated buffers. The buffer size of TFLM is smaller than that of TVM, because TFLM buffers are only used for intermediate data, whereas TVM buffers also contain copies of the model data.

\begin{table}[t]
\small
\centering
\caption{Models for the evaluation.}
\begin{tabular}{|c|c|c|c|}
\hline 
{\bf Name} & {\bf Model size} & {\bf TVM buffer size} & {\bf TFLM buffer size} \\ \hline \hline
MBNET & 17MB & 30MB & 5MB \\ \hline
RSNET & 170MB & 205MB & 24MB \\ \hline
DSNET & 44MB & 55MB & 12MB \\ \hline
\end{tabular}
\label{tab:models}
\end{table}

\begin{figure*}[t]
  \centering
  \begin{minipage}[b]{0.495\textwidth}
    \subfloat[TVM]{\includegraphics[width=0.5\textwidth]{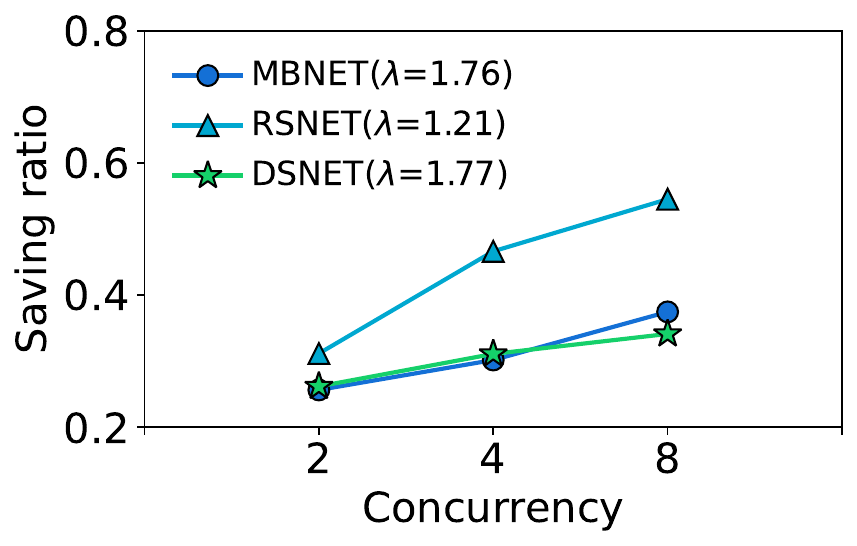}}
    \subfloat[TFLM]{\includegraphics[width=0.5\textwidth]{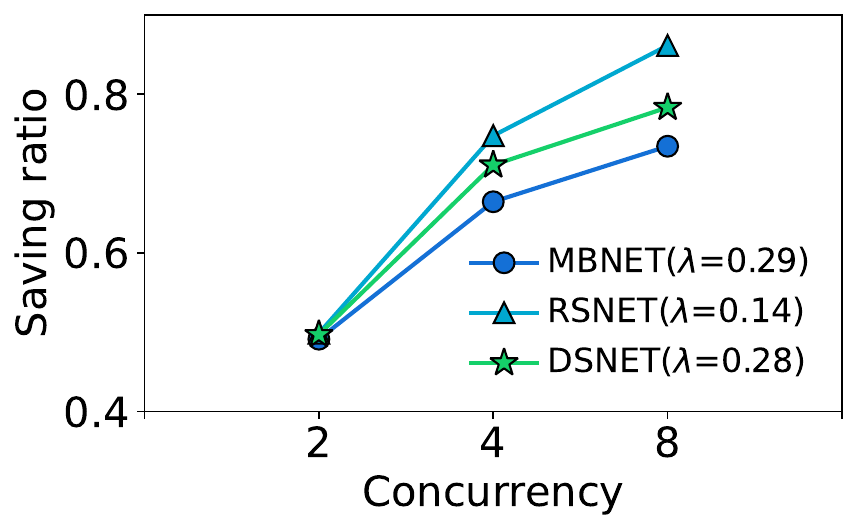}}
    \caption{Enclave memory saving. ($\lambda$: runtime buffer size / model size)}
    \label{fig:mem-save}
  \end{minipage}
  \begin{minipage}[b]{0.495\textwidth}
      \subfloat[SGX2\label{fig:warm-con-sgx2}]{\includegraphics[width=0.5\textwidth]{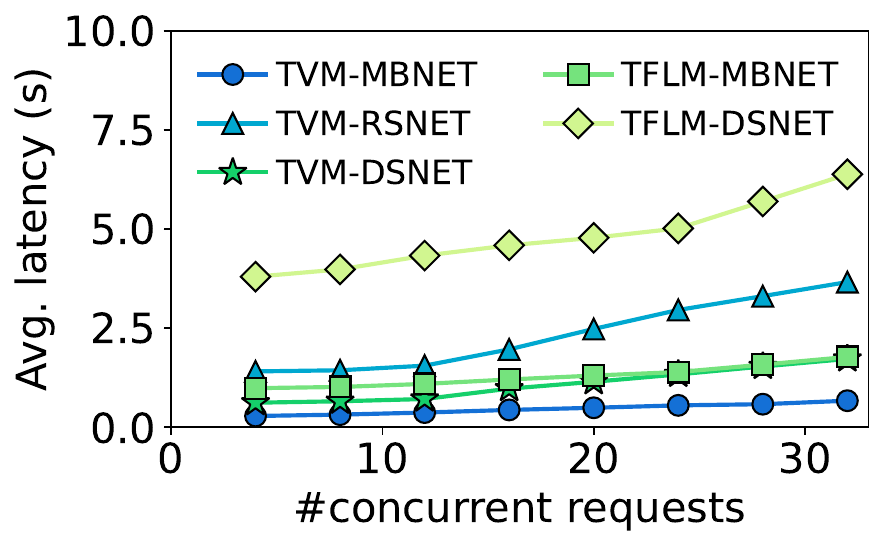}}
      \subfloat[MBNET(SGX1)\label{fig:warm-con-sgx1-mbnet}]{\includegraphics[width=0.5\textwidth]{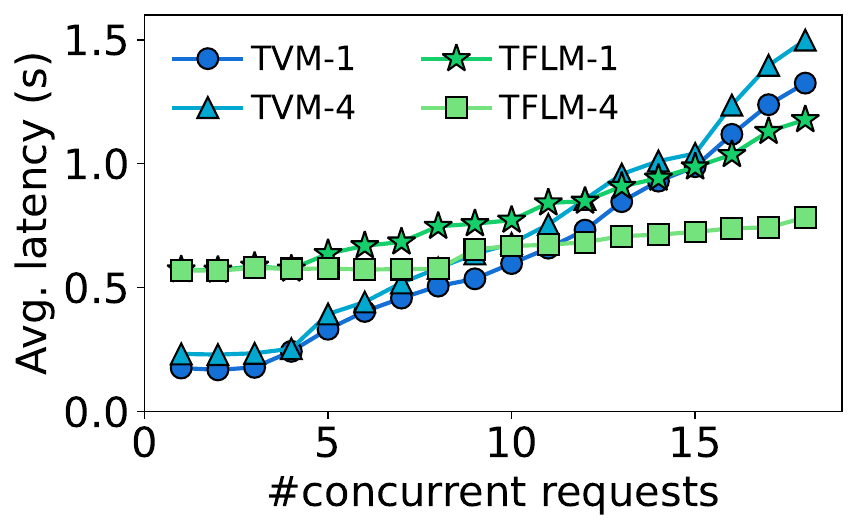}}
      \caption{Latency w.r.t. number of concurrent execution.}
      \label{fig:warm-con-line}
  \end{minipage}
    \vspace{-0.7cm}
\end{figure*}

\begin{figure*}[t]
  \centering
  \subfloat[TVM-MBNET\label{fig:warm-lat-sgx2-mbnet}]{\includegraphics[width=0.25\textwidth]{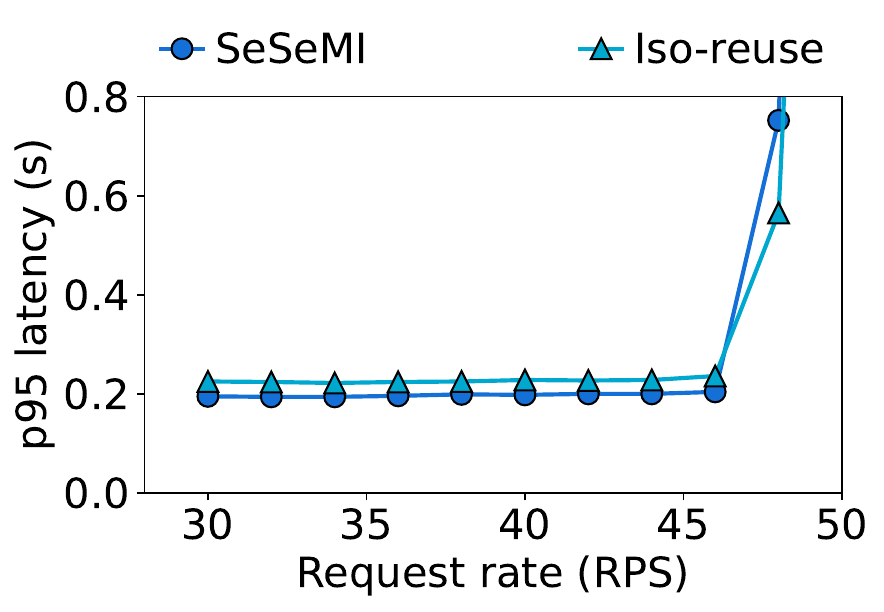}}
  \subfloat[TVM-RSNET\label{fig:warm-lat-sgx2-rsnet}]{\includegraphics[width=0.24\textwidth]{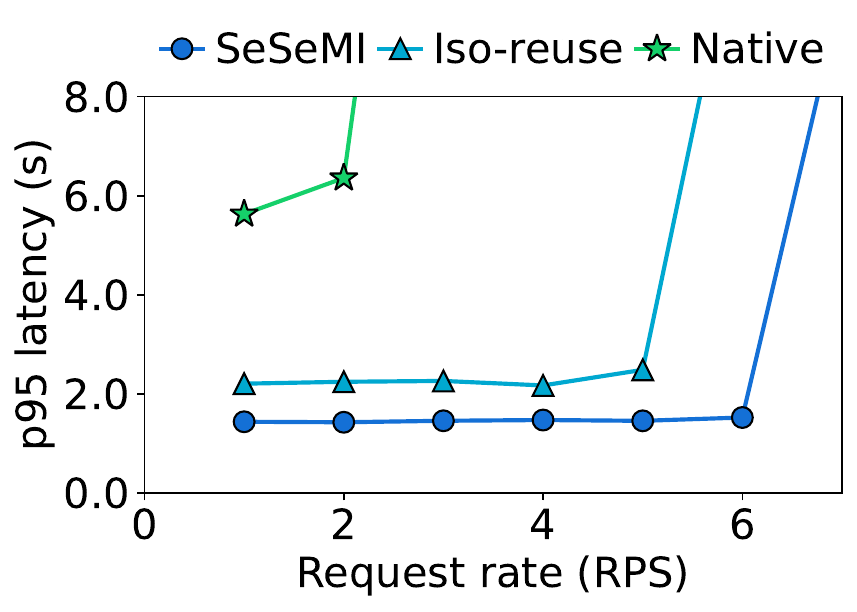}}
  \subfloat[TVM-MBNET(SGX1)\label{fig:warm-lat-sgx1-tvm-mbnet}]{\includegraphics[width=0.24\textwidth]{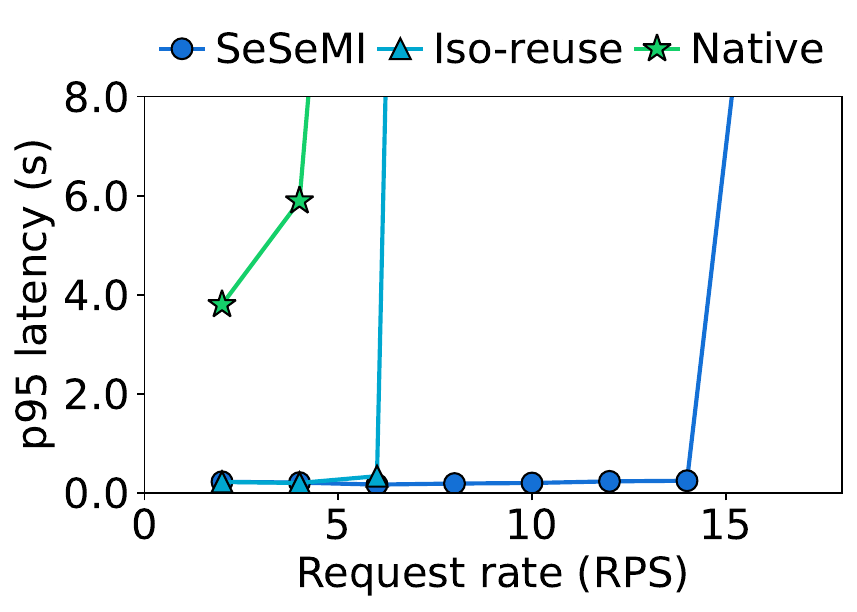}}
  \subfloat[TFLM-MBNET(SGX1)\label{fig:warm-lat-sgx1-tflm-mbnet}]{\includegraphics[width=0.24\textwidth]{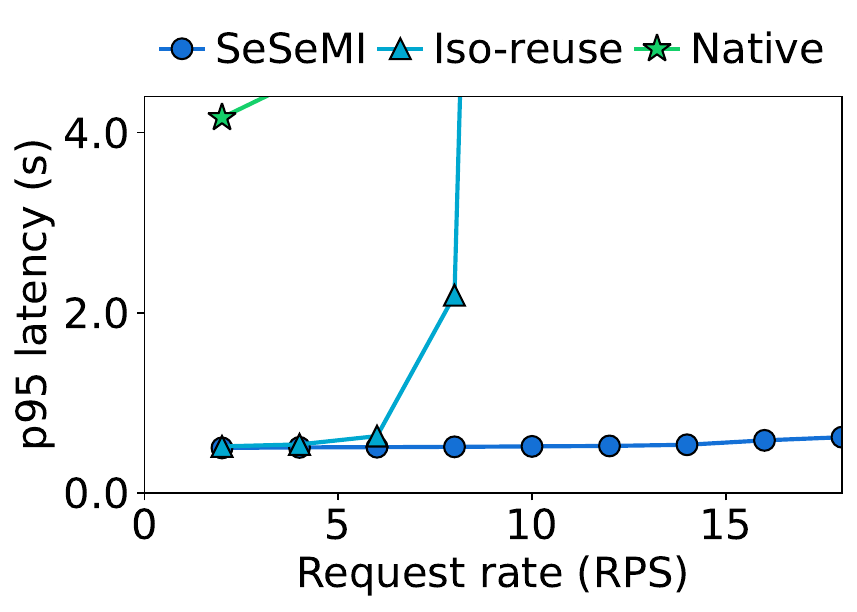}}
  \caption{MBNET and RSNET serving with hot invocations.}
  \label{fig:warm-lat}
    \vspace{-0.3cm}
\end{figure*}

\noindent\textbf{Warm and hot invocation speedup.}
Figure~\ref{fig:ratioperf} breaks down the cold invocation latency of each request over the major serving stages managed by \rtname{}. We can observe that enclave initialization and key fetching that includes remote attestation contribute to a significant portion of latency, over 60$\%$ for TVM models. Another observation is that the runtime initialization stage incurs a minor cost for TFLM models; therefore, reducing this stage will not be as beneficial as that for TVM models.

Figure~\ref{fig:sumperf} compares the execution time for different invocation paths. The sandbox initialization time is not included here because it is independent of the model and framework.
A hot invocation in \rtname{} skips most of the steps in Figure~\ref{fig:serve-stages}. It only executes the model on the request and encrypts the output. A warm invocation includes additional model loading and initialization costs.  A cold invocation contains the costs of all stages. 
It can be seen that the cost is comparable between the warm path and the untrusted execution, and between hot invocation and untrusted execution with model cached. This is expected as model execution is a compute-intensive task.

The advantage of hot invocations in \rtname{} is significant for TVM models that optimize for inference time. For example, for the MBNET model running with TVM, a hot invocation can achieve up to $21\times$ speedup over a cold invocation, whereas a warm invocation achieves a $11\times$ speedup.
The \textit{Native} baseline does not reuse the enclave, which means all the stages in the cold path are involved. The \textit{Iso-reuse} baseline can save the enclave initialization and key retrieval time, but not the model loading and runtime initialization cost. In TVM, the runtime initialization adds 39.6$\%$, 21.3$\%$, and 15.0$\%$ latency compared to the model execution stage for the three models. The model loading cost can be significant for cloud storage. For example, when the models are stored on Azure Blob Storage,
downloading the MBNET, DSNET, and RSNET models in the same region takes around 180$ms$, 360$ms$, and 2100$ms$, respectively. In contrast, the hot invocation in \rtname{} only downloads the model once in the first invocation for each instance.

\noindent\textbf{Memory saving with parallel execution.}  \rtname{} enables one enclave to serve multiple requests. This helps reduce the peak memory consumption because the system needs not to load one copy of the model to serve each request. 
To evaluate the benefit of \rtname{} in saving memory, 
we use the Enclave Memory Measurement Tool for Intel SGX to measure peak memory usage. The results for the two inference frameworks and different concurrency levels are shown in Figure~\ref{fig:mem-save}.
We observe that TFLM's memory saving is more significant, since it only requires buffer space for intermediate data. 
The highest saving is 86.2\% for RSNET under TFLM with 8 threads inside the enclave. 
A higher concurrency level leads to higher memory saving, but it increases the enclave initialization time for a newly launched sandbox, leading to higher latency for cold invocations.

\begin{table}[t]
\small
\centering
\caption{Overhead of stronger isolation on hot invocations.}
\begin{tabular}{|c|c|c|c|}
\hline 
{\bf Name} & {\bf TVM-MBNET} & {\bf TVM-RSNET} & {\bf TVM-DSNET} \\ \hline \hline
Without & 65.79$ms$ & 982.96$ms$ & 388.81$ms$ \\ \hline
With & 268.36$ms$ & 1265.00$ms$ & 587.79$ms$ \\ \hline
\end{tabular}
\label{tab:seq-proc}
\end{table}

\noindent\textbf{Sequential request processing.} As discussed in Section~\ref{sec:implementations}, users can also ensure stronger isolation by restricting single model inference, disabling key caching, enforcing sequential request processing, and clearing the model runtime buffer for each request. This allows the enclave to return to a state with only the loaded model after each invocation, which is similar to other techniques~\cite{Ryoan, Groundhog}. 
Such settings are part of the enclave code and enforced by enclave identity via \ksname{}. 
However, it degrades performance, as hot invocations cannot reuse the cached keys and model runtime, and there is no memory saving when multi-threading is disabled. 
Table~\ref{tab:seq-proc} shows the performance impact. It can be seen that the hot invocation latency increases significantly, by as much as $4\times$. 
In the subsequent experiments, this setting is turned off.

\subsection{Single Node Evaluation}\label{subsec:single-node}

We use one node to launch serverless sandboxes and compare the performance of \name{} against baselines.
We warm up the sandbox instances to ensure there is no cold invocation. 
We send requests to the sandboxes at a fixed rate, and each request contains the same user and model. 

Figures~\ref{fig:warm-lat-sgx2-mbnet} and ~\ref{fig:warm-lat-sgx2-rsnet} show the performance of MBNET and RSNET, respectively. The \textit{Native} baseline has the lowest throughput and highest latency. We do not show \textit{Native} in Figure~\ref{fig:warm-lat-sgx2-mbnet} because it reaches the limit when the request rate is less than 15 requests per second (rps).
For MBNET, the difference between \textit{Iso-reuse} and \name{} is small, and they have roughly the same throughput at 46 rps. TVM-RSNET has a higher runtime initialization cost, so \textit{Iso-reuse} has a lower peak throughput than that of \name{}, because  this baseline repeats the model loading and runtime initialization steps for each request. 

Next, we evaluate the performance overhead of serving concurrent requests.  Figure~\ref{fig:warm-con-sgx2} illustrates how latency increases with more concurrent requests. 
We note that the cost of hot invocation increases despite the memory consumption being below the EPC limit. This is because the model execution is CPU bound. For TVM-RSNET and TVM-DSNET, the latency increases faster when the number of concurrent requests exceeds the number of physical cores (12).

\begin{figure*}[t]
  \centering
  \subfloat[workload]{\includegraphics[width=0.333\textwidth]{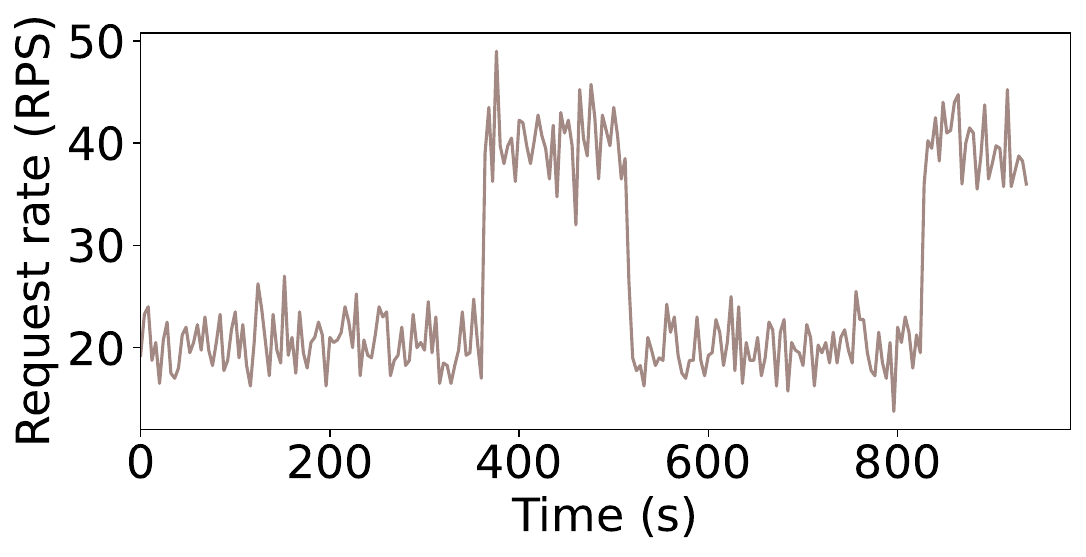}}
  \subfloat[TVM-DSNET\label{multi-node-dsnet}]{\includegraphics[width=0.333\textwidth]{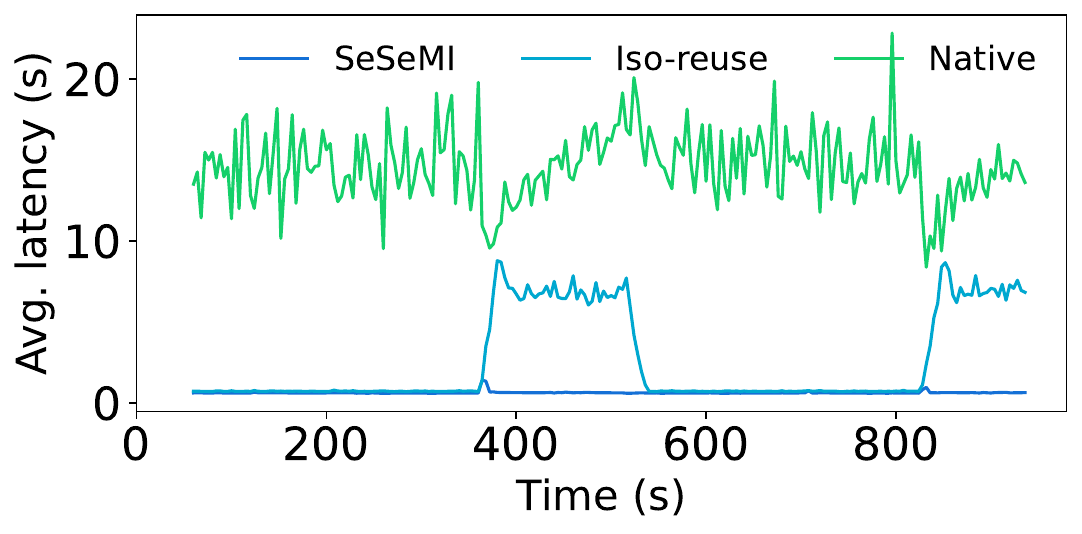}}
  \subfloat[TVM-RSNET\label{multi-node-rsnet}]{\includegraphics[width=0.333\textwidth]{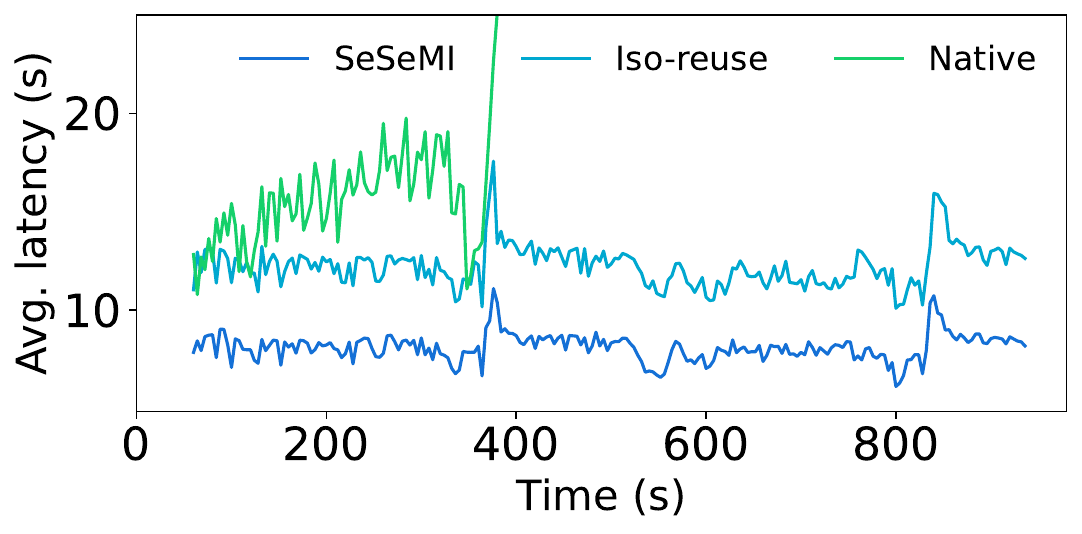}}
  \caption{Latency of serving MBNET under MMPP workload.}
  \label{fig:mmpp-lat}
\vspace{-0.3cm}
\end{figure*}

\begin{figure*}[t]
  \subfloat[TVM-DSNET-1]{\includegraphics[width=0.25\textwidth]{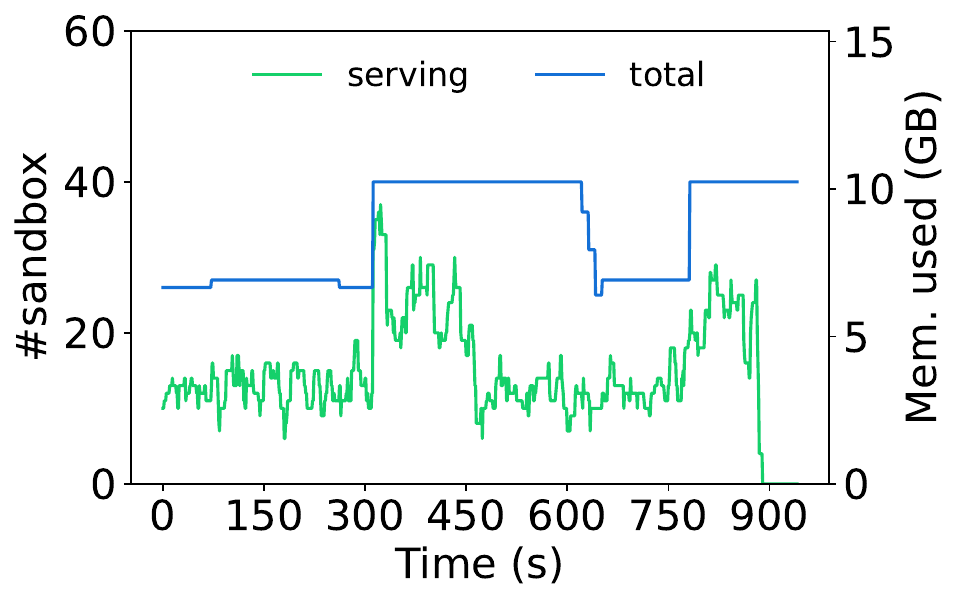}}
  \subfloat[TVM-DSNET-4]{\includegraphics[width=0.25\textwidth]{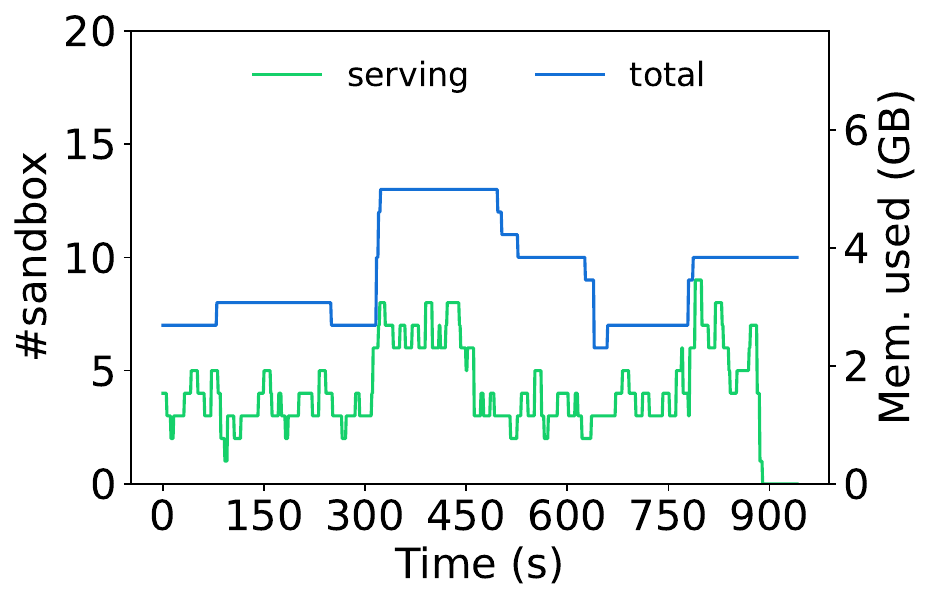}}
  \subfloat[TVM-RSNET-1]{\includegraphics[width=0.25\textwidth]{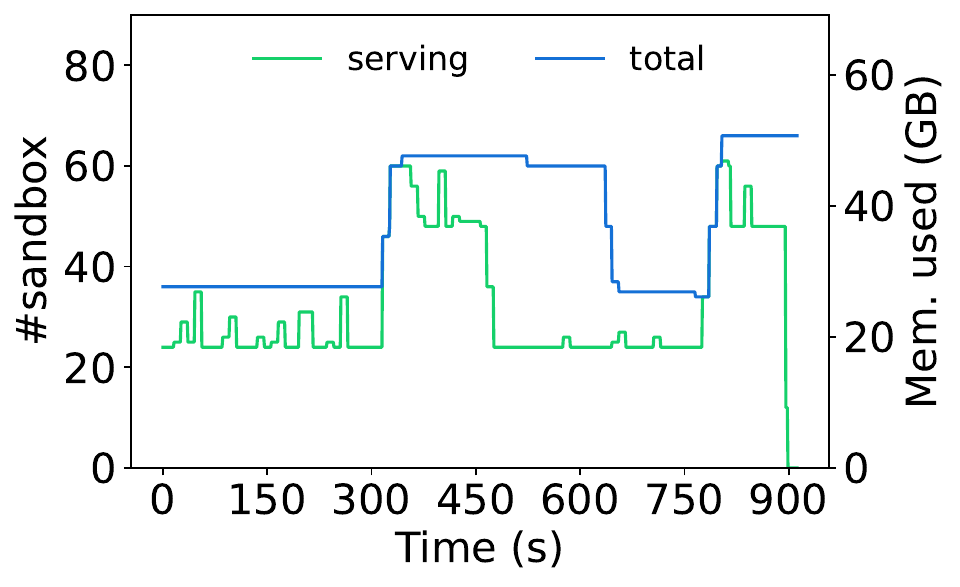}}
  \subfloat[TVM-RSNET-4]{\includegraphics[width=0.25\textwidth]{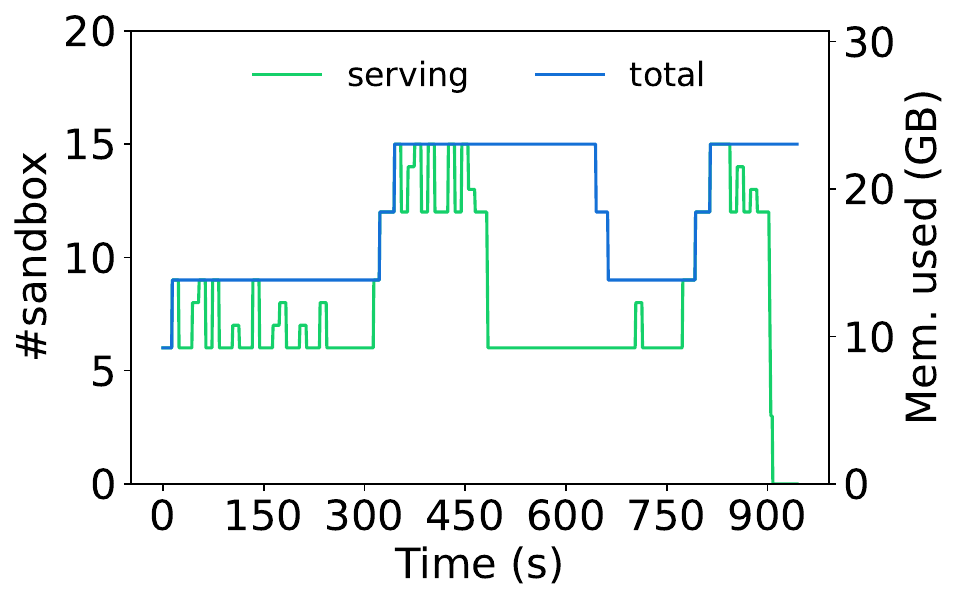}}
  \caption{Memory usage for serving under MMPP workload.}
  \vspace{-0.3cm}
  \label{fig:mmpp-cost}
\end{figure*}

Figure~\ref{fig:warm-con-sgx1-mbnet} shows the overhead when EPC is the bottleneck. The latency starts to increase as the total enclave memory exceeds the EPC limits. TVM reaches the limit before TFLM, because of its higher memory consumption. Figure~\ref{fig:warm-con-sgx1-mbnet} further shows the difference between increasing the number of threads within a single enclave and launching multiple sandbox instances. We compare the latency when running  \rtname{} with one thread (TVM/TFLM-1) and with four threads (TVM/TFLM-4). For TVM, the internal scheduling increases the latency slightly. Although the peak memory usage is smaller in TVM-4, the number of enclave memory pages accessed during model execution is the same as TVM-1, because each thread accesses its own runtime buffer, and the model buffer is not used after the runtime initialization. On the contrary, TFLM runtime buffer is smaller and the multiple threads can access the same memory pages containing the model during model execution. The latency of TFLM-4, therefore, grows at a much slower rate than TFLM-1. The throughput is affected as shown in Figure~\ref{fig:warm-lat-sgx1-tvm-mbnet} and ~\ref{fig:warm-lat-sgx1-tflm-mbnet}, where
TFLM-MBNET can serve more than $18$ rps, whereas TVM-MBNET is limited to $14$ rps.
 
In summary, \name{} outperforms the two baselines because its hot invocations reduce the overall latency to only the cost of model execution. When EPC is not a limiting factor, model inference on a single node becomes compute-bounded, and it is important to reduce the computation of each request as much as possible. When EPC is comparable to the model size, using a framework that optimizes memory consumption can lead to better performance.

\subsection{Multi-Node Evaluation} \label{subsec:multi-node}
We compare \name{} against the two baselines in a multi-node cluster, using 8 nodes for running sandbox instances. 
Because OpenWhisk only uses memory for its scheduling policy and preferably launches instances of a function on the same machine, we configure the invoker memory such that the total number of enclave threads on a node never exceeds the number of physical cores. 
This helps avoid the overhead due to concurrent execution discussed in the previous section.  
We use a Markov-modulated Poisson process (MMPP) workload, which is popular for evaluating serverless model inference~\cite{BATCH-serve, our-work, mark}. The workload alternates the mean request rates between 20 rps and 40 rps. We warm up the system with the workload of 20 rps before collecting the data.

Figure~\ref{fig:mmpp-lat} shows that the \textit{Native} baseline has the highest latencies. Under TVM-DSNET, due to contention, each request takes more than 10$s$ to process on average, and the performance varies significantly as the serverless platform schedules a large number of pending requests. Under TFLM-RSNET, which is a larger model, the serverless platform even becomes unavailable when the request rate suddenly increases.
\textit{Iso-reuse} achieves an average latency of 3.35$s$ for DSNET, while \name{} achieves an average latency of 0.64$s$, an 81$\%$ improvement. For RSNET, the average latency is 12.54$s$ and 8.28$s$ for  \textit{Iso-reuse} and \name{} respectively, due to high contention. It can be seen that  \textit{Iso-reuse} is more sensitive to the sudden increase in request rate. In Figure~\ref{multi-node-dsnet}, \name{} quickly restores to the low latency after the burst of requests.  In contrast, the latency of \textit{Iso-reuse} increases significantly and remains high for a long period after the request burst. 

To evaluate how concurrent execution of \rtname{} affects the cost, we run the same workload with one thread (TVM-1) and with four threads (TVM-4) per enclave. Each thread in an enclave has its own runtime buffer, but they share the same model buffer within an enclave. The memory consumption is measured by the number of sandbox instances times the memory budget configured for the function. The results are shown in Figure~\ref{fig:mmpp-cost}.
We compute the cost as the integral of enclave memory consumption over the workload duration, a commonly adopted method on serverless platforms. The memory budget is 256MB for TVM-DSNET-1, 384MB for TVM-DSNET-4, 768MB for TVM-RSNET-1, and 1536MB for TVM-RSNET-4. 
For DSNET, the memory consumption of TVM-4 is 1459GB-Second compared to 3543GB-Second of TVM-1, a cost reduction of 59$\%$.
For RSNET, TVM-4 reduces the cost by 48 $\%$ from 2273GB-Second to 1179GB-Second.

\subsection{\fpname{} Evaluation} \label{subsec:fnpacker-evaluation}
To evaluate the effectiveness of \fpname{} on infrequent and unpredictable workloads for multi-model serving, we generate a workload by mixing two representative workload patterns in MLPerf~\cite{mlperf}. One is based on a  Poisson arrival simulating access to popular models ($m_0$ and $m_1$), each receiving requests at 2 rps for 8 minutes. The other consists of two interactive sessions at around 4 and 6 mins, in which a set of models ($m_0 - m_4$) are sequentially queried, representing the scenario that a model user tries out multiple models for his sample data. We denote TVM-RSNET models with different IDs as $m_0 - m_4$. 

We compare \fpname{} with two baseline solutions described in Section~\ref{subsec:\fpname{}}: \textit{One-to-one} that creates one function endpoint for each model, and \textit{All-in-one} that creates one function endpoint to serve all the models. 
Table~\ref{tab:fnpack-m0} shows the average latency of requests from the Poisson arrivals targeting $m_0$ and $m_1$. 
Since \fpname{} adaptively schedules the requests, $m_0$ and $m_1$ are assigned to two different and exclusive endpoints, which results in no interference when serving requests.
Therefore, we observe no impact on the interactive sessions nor between Poisson request streams for $m_0$ and $m_1$. In contrast, in \textit{All-in-one}, the requests to $m_0$ and $m_1$ interfere with each other and cause the sandbox instances repeatedly swap the serving model. The average latency is increased by over 16$\%$ compared to the two baselines.

Table~\ref{tab:fnpack-interactive} shows the latency of executing model $m_0-m_4$ during the interactive session. 
The latency of \textit{One-to-one} and \fpname{} are similar because they reuse instances serving the continuous traffic. The latency is higher in \textit{All-in-one}, because of model switching. 
The \textit{One-to-one} baseline incurs a cold start for each of $m_2$, $m_3$, and $m_4$ for the first session, which leads to significant overhead. In the second session, requests reuse the sandboxes launched during session 1. \fpname{} launches a sandbox instance under a function endpoint different from that of $m_0$ and $m_1$.  For subsequent queries, it detects that the endpoint is idle and routes the requests to it. Thus, only the first request to $m_2$ is served by the cold invocation, and the rest are served by warm invocations.
If the access is through remote cloud storage, the model switching cost will be even higher. In that case, interference in \textit{All-in-one} is more severe, which may cause the system to launch additional sandboxes to serve requests. 
By reducing the function execution time, \fpname{} helps lower the model owner's financial cost for managing the service.

\begin{table}[t]
\small
\centering
\caption{Latency of models with Poisson traffic.}
\vspace{-2mm}
\begin{tabular}{|c|c|c|c|}
\hline 
{} & {\bf All-in-one} & {\bf One-to-one} & {\bf FnPacker} \\ \hline \hline
Avg. latency (ms) & 1700.50 & 1456.01 & 1465.79 \\ \hline
\end{tabular}
\label{tab:fnpack-m0}
\end{table}

\begin{table}[t]
\small
\centering
\caption{Latency of serving interactive queries.}
\vspace{-2mm}
\begin{tabular}{|c|c|c|c|c|}
\hline 
{Session} & {Model} & {\bf All-in-one} & {\bf One-to-one} & {\bf FnPacker} \\ \hline \hline
\multirow{5}{*}{\shortstack[c]{Session 1 \\ (ms)}} & $m_0$ & 1513 & 1399 & 1464 \\ \cline{2-5}
                            & $m_1$ & 1805 & 1438 & 1480 \\ \cline{2-5}
                            & $m_2$ & 2045 & 9405 & 10166 \\ \cline{2-5}
                            & $m_3$ & 3556 & 9752 & 2008 \\ \cline{2-5}
                            & $m_4$ & 1907 & 9923 & 2045 \\ \hline
\multirow{5}{*}{\shortstack[c]{Session 2 \\ (ms)}} & $m_0$ & 1548 & 1409 & 1434 \\ \cline{2-5}
                            & $m_1$ & 1782 & 1306 & 1416 \\ \cline{2-5}
                            & $m_2$ & 2038 & 1409 & 1807 \\ \cline{2-5}
                            & $m_3$ & 1974 & 1400 & 1546 \\ \cline{2-5}
                            & $m_4$ & 1989 & 1385 & 1799 \\ \hline
\end{tabular}
\label{tab:fnpack-interactive}
\end{table}

\section{Related Work}\label{sec:related-work}

\noindent
\textbf{Secure data management using TEEs.}
Trusted hardware is an effective primitive for building secure data systems in untrusted environments. ~\cite{operon, azure-sql-ae, stealthdb, enclavedb, ironsafe, veridb} runs the database engines inside TEEs to protect SQL execution over sensitive data. ~\cite{vc3, opaque, seccask, tgcb, data-station, ccf} targets general data analytic in non-serverless settings. ~\cite{Chiron, SecureTF, Vessels, Occlumency}
use TEEs for model inference, and are the most related to our work. However, they are designed to run on reserved instances (as opposed to serverless). Occlumency~\cite{Occlumency} and Vessels~\cite{Vessels}
propose techniques to reduce memory overheads in SGX1. These techniques can be integrated into \name{}. However, we show that for SGX2 the performance bottleneck has shifted from memory to CPU. 
\name{} makes use of TEE to protect
the users’ data and the model owner’s model trained on sensitive
information and prevents them from unauthorized access, while
allowing the cloud to elastically scale out. 

\noindent
\textbf{Serverless computing with secure hardware.} 
~\cite{s-faas, securev8, clemmys, reusable-enclave} requires extensive modification to the serverless platform control plane.
S-FaaS~\cite{s-faas} proposes a Key Distributed Enclave (KDE),
but it generates and distributes the same key for a 
function, which cannot provide fine-grained access control over users or data required for serving models. 
Moreover, we demonstrate that they suffer long latency and poor scalability. 
PIE~\cite{PIE} and PENGLAI~\cite{Penglai} propose new enclave initialization mechanisms to improve the scalability of enclave launching in serverless computing, but they require hardware support and are incompatible with existing serverless platforms.
Faastlane~\cite{faastlane} assumes an honest cloud provider and uses Intel Memory Protection Key to isolate sensitive data processing. 
\name{} assumes the cloud is untrusted, requires no changes to cloud providers' infrastructure, and optimizes for model inference. 

\noindent
\textbf{Reusing states in serverless sandbox instances.}~\cite{fireworks, seuss, replayable-execution, sock, sand} aims to speed up the initialization of runtime and libraries for conventional serverless computing without TEE. 
OpenWhisk~\cite{openwhisk}, Knative~\cite{knative} and Cloud Function v2~\cite{gcp-cf2} supports scheduling multiple requests to a serverless instance, which can be leveraged by ~\rtname{}.
SAND~\cite{sand}, Fifer~\cite{Fifer}, and Nightcore~\cite{nightcore} allow concurrent execution of functions in a workflow inside a sandbox instance. 
\rtname{} is designed for secure model inference, 
in which we reuse the enclave and loaded models to reduce latency, memory, and thus costs.

\section{Conclusions}\label{sec:conclusion}

In this paper, we propose a novel secure serverless model inference system \name{}. It enables model inference on sensitive data and protects model confidentiality in an untrusted cloud environment leveraging trusted hardware. It consists of three non-intrusive components to existing serverless platforms, namely a trust delegation service \ksname{}, an efficient enclave runtime \rtname{}, and a flexible packing module \fpname{}. We conduct extensive experiments using three deep learning models under two inference frameworks and on two versions of SGX, demonstrating that \name{} achieves high performance and is cost-effective compared to two baselines.

\bibliographystyle{IEEEtran}
\balance
\bibliography{ref}
\clearpage
\appendix
\section{Appendix}

\subsection{Remote Attestation in \name{}} 
\label{appendix:ra}

In \name{}, remote attestation is implemented with standard TLS to set up a secure channel for transmitting sensitive information from the client to \ksname{} enclave and between \ksname{} enclave and \rtname{} enclave. We refer to ~\cite{ratls} and its ~\cite{ratls-code} for our implementation. The TLS certificate embeds the attestation report. The report binds the enclave identity, includes the SGX platform information, and is signed by a key that Intel securely provision to the SGX platform. The receiver of the certificate can verify with Intel (i.e. through IAS for EPID or the ECDSA signature) the authenticity of the attestation report and then check the enclave identity with the expected value. For the mutual attestation we implement between \ksname{} enclave and \rtname{} enclave, the TLS certificate verification happens inside the enclave to protect their execution integrity and secure channel established. The identity of \rtname{} enclave is stored inside \ksname{} by the model owner and end users beforehand. The \ksname{} enclave identity can be compiled with the \rtname{} enclave codes, so tampering with it will change the \rtname{} enclave value, and the \rtname{} enclave will fail the access control checking.

\subsection{Code and Data Protected by Enclave Identity} 
\label{appendix:mrenclave}

Here, we explain the enclave identity that is crucial in \name{} to identify the codes and data inside an enclave, ensuring their correctness and protecting execution integrity. The enclave identity is computed with a secure hash function during enclave initialization over the input and process of setting up the enclave. The value generated is stored inside the enclave and added to the report during remote attestation. Hence, its value cannot be tampered with. The enclave identity is the same for the given compiled codes and data, independent of the server it runs on. Therefore identity checking is unaffected by the function scheduling over multiple servers in the serverless platform. For \ksname{}, the functions and data structure described in \ref{alg:key-service} are part of the enclave codes, and their execution is guaranteed to be correct by the TEE. For \rtname{}, codes implementing the two functions in \ref{alg:executor}, the data structure to synchronize caching and access to key pairs and model content inside the enclave, and the variables discussed in Section~\ref{sec:implementations} to enforce additional execution restrictions are part of the enclave codes. The input content, keys, and model content are processed inside the enclave but do not constitute the enclave identity. With encryption, when those data are tampered with, it will trigger execution errors in the enclave functions and be identified. 

\subsection{Enclave Launch Overhead of \rtname{}}

\begin{figure}[ht]
  \centering
  \subfloat[SGX2 \label{fig:launch-con-sgx2}]{\includegraphics[width=0.23\textwidth]{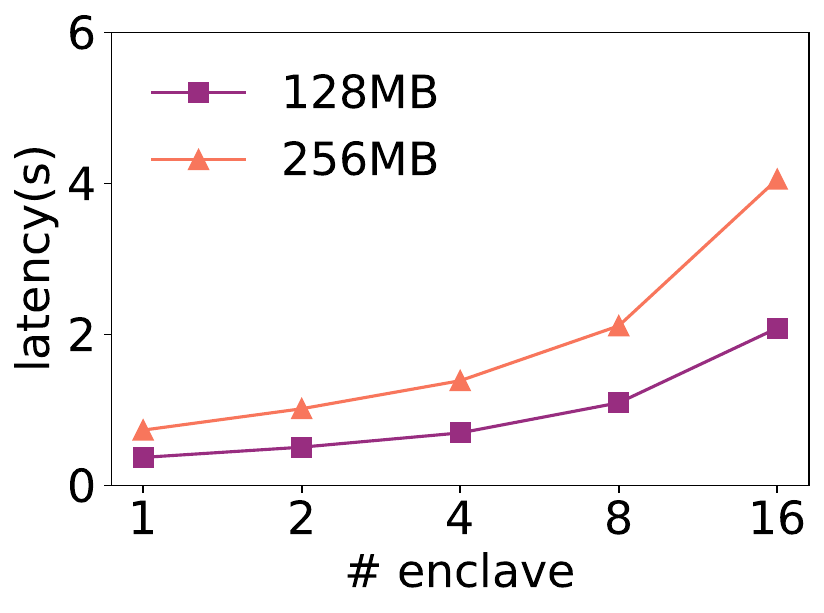}}
  \subfloat[SGX1 \label{fig:launch-con-sgx1}]{\includegraphics[width=0.24\textwidth]{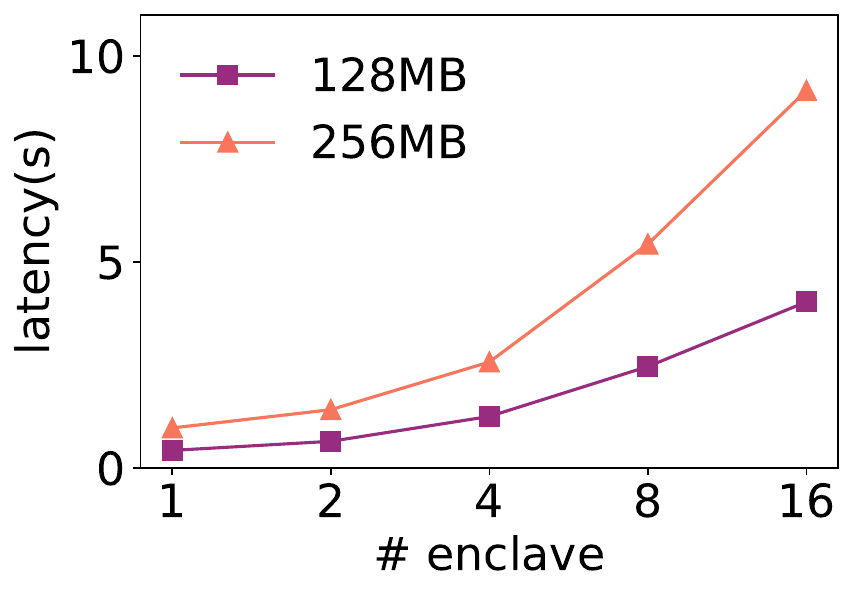}}
  \caption{Enclave initialization overhead.}
  \label{fig:enclave-init-overhead}
\end{figure}

\begin{figure}[ht]
  \centering
  \subfloat[SGX2-ECDSA \label{fig:ra-con-sgx2}]{\includegraphics[width=0.23\textwidth]{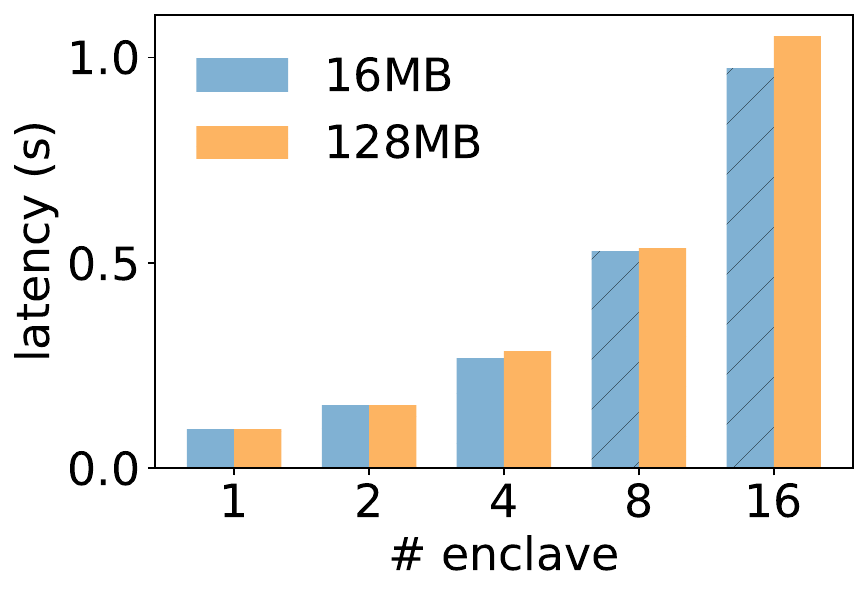}}
  \subfloat[SGX1-EPID \label{fig:ra-con-sgx1}]{\includegraphics[width=0.22\textwidth]{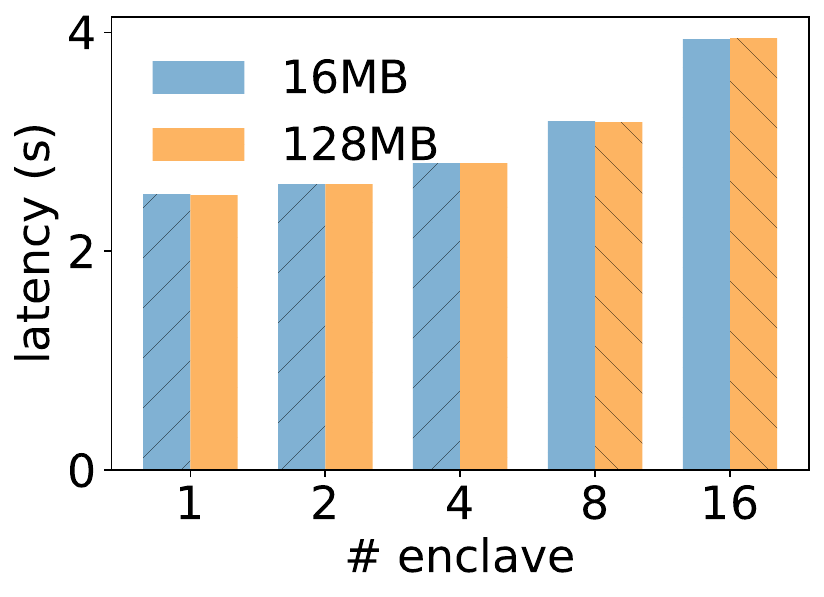}}
  \caption{Remote attestation overhead.}
  \label{fig:ra-overhead}
\end{figure}

Figures~\ref{fig:launch-con-sgx2} and \ref{fig:ra-con-sgx2} show the cost of SGX2 enclave
initialization and remote attestation (which is necessary in the key retrieval step). 
It can be seen that the enclave initialization time is higher for a larger enclave memory size, and it increases with the number of enclaves being launched concurrently. 
When starting multiple enclaves at the same time,
the average initialization time increases. For example, with 16 concurrent enclaves of 256MB, 
each takes 4.06$s$ on average to complete initialization.
This cost restricts the scalability of serverless platforms to handle concurrent requests on the same physical machine. 
The cost of remote attestation (RA) is high, but it is independent of the enclave size. 
We observe an increase in latency when multiple enclaves generate quotes that are used to prove the identity of enclaves in the remote attestation. For example, the average latency increases from less than 0.1$s$ for 1 enclave to 1$s$ when 16 concurrent enclaves are running quote generation.

Similar trends are observed on SGX1 platforms as shown in Figure~\ref{fig:launch-con-sgx1} and ~\ref{fig:ra-con-sgx1}. The enclave initialization has higher latency when multiple enclaves are concurrently launched because all memory reserved for enclaves need to be added to the Enclave Page Cache (EPC) which is limited to up to 128MB on an SGX1 machine. On SGX2 machines, EPC can be configured to 64 GB. The attestation based on Intel Enhanced Privacy ID (EPID)~\footnote{https://www.intel.com/content/www/us/en/developer/articles/technical/intel-enhanced-privacy-id-epid-security-technology.html} 
available on SGX1 also takes longer than the ECDSA-based attestation Intel SGX Data Center Attestation Primitives (DCAP)~\footnote{https://www.intel.com/content/www/us/en/developer/articles/technical/quote-verification-attestation-with-intel-sgx-dcap.html}
, because it requires communication with Intel Attestation Service.

\subsection{Memory Overhead of TEE Protection}
\label{appendix:memory}

The input data and model are encrypted under our attack model when transmitted outside of enclaves. The encrypted contents are copied into the enclave and followed by decryption inside. Therefore, the enclave memory needs to have additional space to hold the encrypted copy of data. It increases the peak memory usage of serverless function execution and thereby the cost. It is another source of overhead when enabling TEE for serverless model serving.  The model is the largest encrypted data being loaded into the enclave. For the evaluation, memory configurations for the various enclave are $0x3000000$ for TFLM-MBNET, $0x16000000$ for TFLM-RSNET, $0x6000000$ for TFLM-DSNET, $0x4000000$ for TVM-MBNET, $0x23000000$ for TVM-RSNET and $0x8000000$ for TVM-DSNET for concurrency setting of 1. Allowing an enclave to support multiple concurrent requests allows amortizing this memory overhead.   

\subsection{Developing Functions for Deployment}
Developers write function codes to implement \textsc{model\_exec}, and, given the model, model runtime, and user request as input parameters, the function codes will be similar to those running a model inference without SGX.
It uses the inference framework API to run inference on a request. The inference codes will be compiled into the enclave codes when the \rtname{} is built.

To extend \rtname{} to support other inference frameworks, the inference framework libraries need to be patched to be used inside SGX enclaves. Thereafter, one only needs to implement \textsc{model\_load}, \textsc{runtime\_init}, and \textsc{prepare\_output} APIs using the inference framework. Then the important model inference stages will be automatically managed by \rtname{}. That is also how TFLM and TVM are integrated with \rtname{}. \textsc{model\_load} embeds \textsc{oc\_load\_model} to load the model, decrypt, and deserialize the model to an object used by the inference framework. \textsc{runtime\_init} is called by a thread that enters the enclave and initializes a model runtime object that will be maintained in the thread context. 
\textsc{prepare\_output} serializes the output from the model runtime instance to a byte sequence. \rtname{} automatically caches the return of \textsc{model\_load} into the model cache and maintains the model runtime instance in the thread context of a thread after it calls \textsc{runtime\_init}. 

\subsection{Additional Evaluation Setup Details} \label{appendix:setup}

\begin{table}[t]
\small
\centering
\caption{Configuration parameters.}
\begin{tabular}{|p{2cm}|p{3.5cm}|p{1.7cm}|}
\hline 
{\bf Name} & {\bf Definition} & {\bf Value} \\ \hline \hline
Invoker memory (SGX2) & Amount of memory per node to launch serverless instances & 1GB - 64GB \newline (default: 64GB) \\ \hline
Invoker memory (SGX1) & Amount of memory per node to launch serverless instances & 12.5GB \\ \hline
Container unused timeout & How long a container is kept warm & 3 minutes  \\ \hline
Container \newline memory budget & Memory limit of a container instance (for each function) & Multiple of \newline 128MB \\ \hline
Enclave \newline concurrency & Number of TCSs for an enclave & 1-8 (default: 1) \\ \hline
\end{tabular}
\label{tab:sw-setup}
\end{table}

Attestation services are set up according to Intel's guidelines for development. For SGX2 nodes, we use Intel SGX DCAP 1.11, which is compatible with Intel SGX SDK 2.14. The Intel PCCS service for remote attestation is set up on the node that is running \ksname{} and \fpname{}. For SGX1 experiments, enclaves communicate with Intel Attestation Service over the Internet to prepare the remote attestation reports. 

To enable access to Intel SGX, the OpenWhisk invokers are configured to add extra flags that mount the SGX driver and aesmd, when launching docker containers. Prometheus is deployed with OpenWhisk to collect the metrics of containers. Table~\ref{tab:sw-setup} lists the OpenWhisk parameters. We configure the OpenWhisk system to keep containers warm for 3 minutes after the last invocation. The memory budget for a container is the smallest multiple of 128MB that is required for a given model. 128MB is a common memory provisioning granularity used by existing cloud service providers. For multi-node evaluations, we configure the memory available to the invoker such that the number of total TCS used by \rtname{} at most is the number of processor cores on a machine, such that OpenWhisk balances the load across multiple machines. 

\subsection{Execution Time Breakdown} \label{appendix:breakdown}

\begin{figure}[t]
  \centering
    \includegraphics[width=0.48\textwidth]{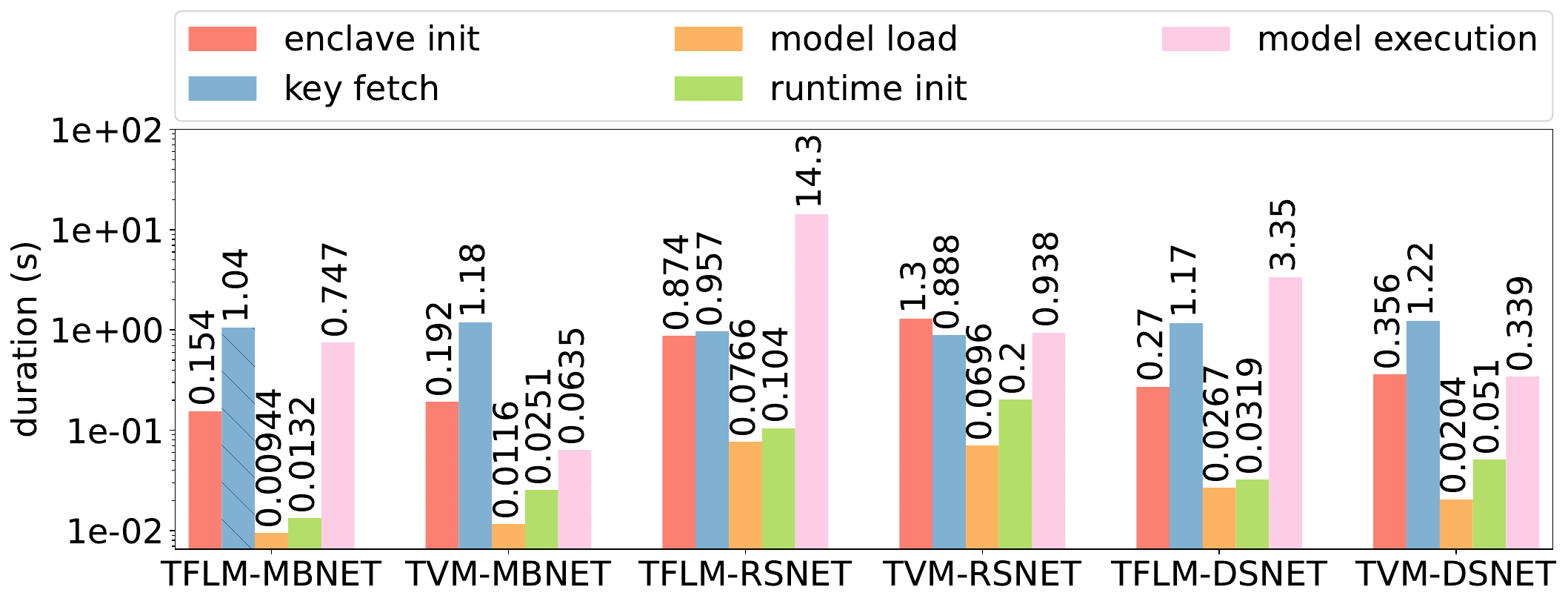}
    \caption{Execution time breakdown for SGX2}
    \label{fig:rawtperf}
\end{figure}

\begin{figure}[t]
  \centering
    \includegraphics[width=0.48\textwidth]{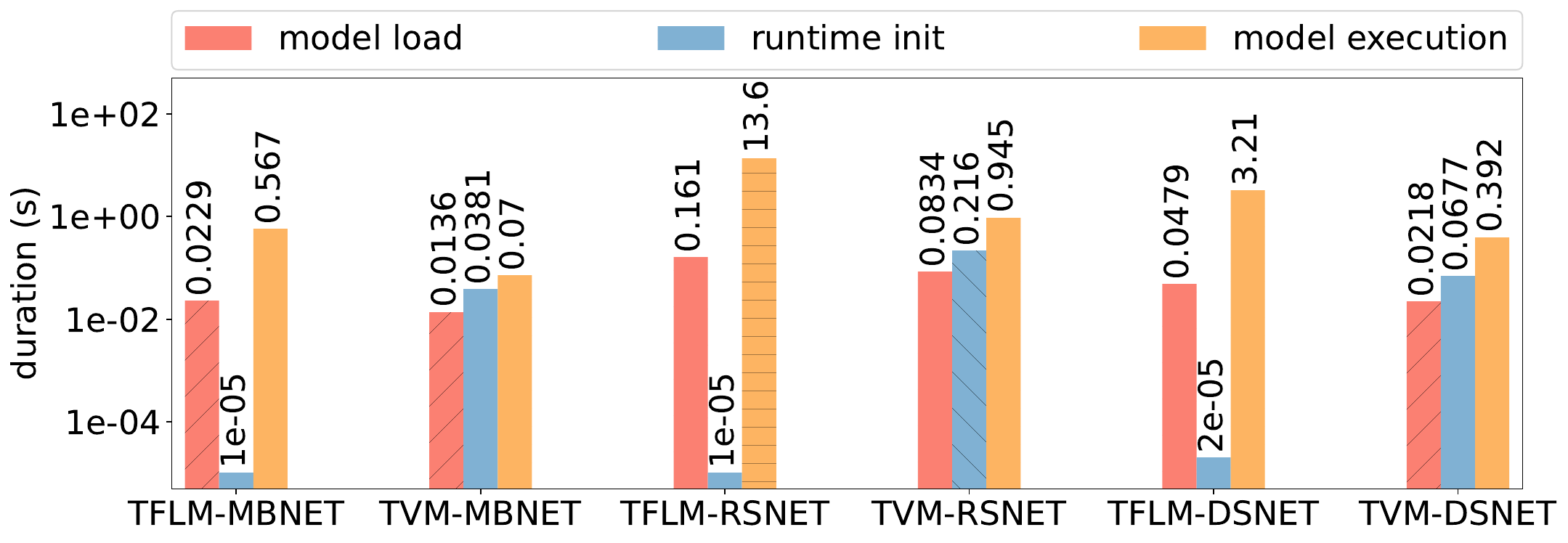}
    \caption{Execution time breakdown outside SGX}
    \label{fig:rawuperf}
\end{figure}

Figure~\ref{fig:rawtperf} and Figures~\ref{fig:rawuperf} show the breakdown of execution time for processing one request with and without SGX on an SGX2 machine.
It can be seen that the overhead primarily comes from enclave initialization and attestation to retrieve the decryption keys. The EPC (64GB for the SGX2 machines used) is much larger than the memory required to run model inference, so the 3 stages in common have minimal performance differences.

\end{document}